\documentclass[twocolumn]{article}
\usepackage{color}
\usepackage{amsmath}
\usepackage{bm}
\usepackage{bbm}
\usepackage{cite}
\usepackage{paralist} 
\usepackage{algorithm}
\usepackage{algorithmic}
\usepackage{graphicx}
\usepackage{float}
\usepackage{color}
\usepackage{amsmath}
\usepackage{cleveref}
\usepackage{bm} 
\usepackage{amssymb} 
\usepackage{geometry}
\usepackage{multicol}  
\usepackage{multirow}  
\usepackage{url}
\usepackage{indentfirst}
\usepackage{authblk}
\usepackage{abstract}
\usepackage{titlesec}

\titlespacing*{\section} {0pt}{1pt}{1pt}
\titlespacing*{\subsection} {0pt}{0.8pt}{0.8pt}
\titlespacing*{\subsubsection}{0pt}{0.7pt}{0.7pt}
\titlespacing*{\paragraph} {0pt}{0.5pt}{0.5pt}
\titlespacing*{\subparagraph} {\parindent}{2pt}{2pt}

\vspace{-0.1cm}  
\title{Special-Purpose Quantum Processor Design}
\author[1,2]{Bin-Han LU}
\author[1,2]{Yu-Chun WU \thanks{Corresponding author: wuyuchun@ustc.edu.cn}}
\author[3]{Wei-Cheng KONG}
\author[1,2]{Qi ZHOU}
\author[1,2,3]{Guo-Ping GUO \thanks{Corresponding author: gpguo@ustc.edu.cn}}
\affil[1]{Key Laboratory of Quantum Information, Chinese Academy of Sciences, School of Physics, University of Science and Technology of China, Hefei, Anhui, 230026, P. R. China}
\affil[2]{CAS Center For Excellence in Quantum Information and Quantum Physics, University of Science and Technology of China, Hefei, Anhui, 230026, P. R. China}
\affil[3]{Origin Quantum Computing Company Limited, Hefei {\rm 230026}, China}
\date{}

\newtheorem{Def}{Definition}

\begin{document}

\twocolumn[
\maketitle
\begin{onecolabstract}
\begin{footnotesize}
Full connectivity of qubits is necessary for most quantum algorithms, which is difficult to directly implement on Noisy Intermediate-Scale Quantum processors. 
However, inserting swap gate to enable the two-qubit gates between uncoupled qubits significantly decreases the computation result fidelity.
To this end, we propose a Special-Purpose Quantum Processor Design method that can design suitable structures for different quantum algorithms.
Our method extends the processor structure from two-dimensional lattice graph to general planar graph and 
arranges the physical couplers according to the two-qubit gate distribution between the logical qubits of the quantum algorithm and the physical constraints. 
Experimental results show that our design methodology, compared with other methods, could reduce the number of extra swap gates per two-qubit gate by at least $104.2\%$ on average. 
Also, our method’s advantage over other methods becomes more obvious as the depth and qubit number increase. 
The result reveals that our method is competitive in improving computation result fidelity and it has the potential to demonstrate quantum advantage under current technical conditions.
\end{footnotesize}
\end{onecolabstract}
]
\section*{\begin{large}INTRODUCTION\end{large}}
\begin{small}
In recent years, quantum computing has progressed to a
“Noisy Intermediate-Scale Quantum(NISQ) era” in which quantum processors have dozens to hundreds of noisy qubits
\cite{Quantumadvantagewithshallowcircuits,Quantumadvantagewithnoisyshallowcircuits,prekill2018}. 
NISQ processors have short coherence time and quantum operations with nonzero error rates \cite{almudever2017}. 
Besides, only a subset of physical qubit pairs are coupled for two-qubit gates, i.e., the Physical Coupling Graphs(PCG) of NISQ processors are not complete. 
Here, the vertexes of the PCG are physical qubits and the edges are physical couplers.
Quantum algorithms are represented by a quantum circuit model. In general, quantum circuits allow two-qubit gates to act on any qubit without restriction.
Therefore, a transformation process that insert extra swap gates before two-qubit gates between uncoupled physical qubits to move the qubits to coupled qubits is required.
Because swap gate consists of imperfect gates supported by NISQ processors, this process will significantly decrease the computation result fidelity \cite{cowtan2019}.

To improve the fidelity, two solutions are proposed,
one is finding the optimized transformation process on the fixed PCG to reduce extra swap gates
\cite{CODAR,zhou2020,niu2020,li2019,pozzi2020,tannu2019,lin2014,shafaei2014,wille2014,lye2015,wille2019,kole2020,siraichi2018,bhattacharjee2019,venturelli2020,murali2019,TacklingtheQubitMappingProblemforNISQ-EraQuantumDevices,tan2020}. 
The other is designing the PCG according to a weighted Circuit Coupling Graph(CCG) of the quantum circuit.
Here the vertex of CCG is logical qubit and the edge weight is the number of successive two-qubit gate blocks \cite{li2020}.
Since the designed PCG is more suitable for the given CCG, the transformation process adds less swap gates than on fixed PCG.
The commonly used PCGs are two-dimensional(2d) lattice graphs because they can be used in Quantum Error Correction(QEC) \cite{fowler2012,devitt2009,jones2010,chamberland2020}
and fabricating these structures on superconducting system is technically practicable \cite{boixo2018,arute2019}.
In \cite{li2020}, the authors proposed a method that designs the PCG based on 2d lattice graph.

However, under current technology, NISQ processor has too few qubits to perform fault-tolerant universal quantum algorithms 
\cite{TacklingtheQubitMappingProblemforNISQ-EraQuantumDevices}.
Therefore, making quantum processor into lattice graphs is not necessary in the NISQ era.
As the general CCG is not a lattice graph, we can extend PCG from lattice graph to general planar graph.
Actually, in \cite{brink2018,almudever2017,AQuantumEngineer'sGuidetoSuperconductingQubits}, we know that manufacturing the general planar PCG is also feasible.

In this paper, we propose a Special-Purpose Quantum Processor Design(SPQPD) method. 
The method extends PCG from lattice-graph to general planar graph. 
We first get the CCG of the circuit, and then arrange the physical couplers according to the CCG under physical constraints. 
Finally, we get a manufacturable PCG. When using realistic algorithms \cite{Gadi2016} as circuit benchmarks, we find that compared with other methods, 
our method, on average, reduces the extra swap gate number by at least $104.2\%$. 
To further verify the effectiveness of SPQPD, we use a series of random circuits as benchmarks and carry out the control variable experiments. 
When fixing qubit number, the number of extra gates in our method decreases as depth increases while other methods either increase or have no obvious trend; 
when fixing algorithm depth, the growth rate of extra gates to the qubit number is reduced at least by $37.1\%$. 
The results reveal that the PCG designed by SPQPD outperforms its competitors. Hence, it has great potential to demonstrate the quantum advantage in the NISQ era.

If quantum processor has enough qubit resources for QEC, i.e., transformation won't reduce the fidelity of the computation results,
SPQPD method can still play an important role because it can reduce the number of swap gates, 
the depth of the circuits, and the time of quantum algorithm execution effectively.

\end{small}
\section*{\begin{large}RESULTS\end{large}}
\subsection*{\begin{normalsize}Technical Constraints of Manufacturable PCG\end{normalsize}}
\begin{small}
There are some technical constraints to NISQ processor's PCG structure.

\begin{figure*}[htbp]
\centering
\begin{minipage}[t]{0.3\textwidth}
\centering
\includegraphics[width=\textwidth]{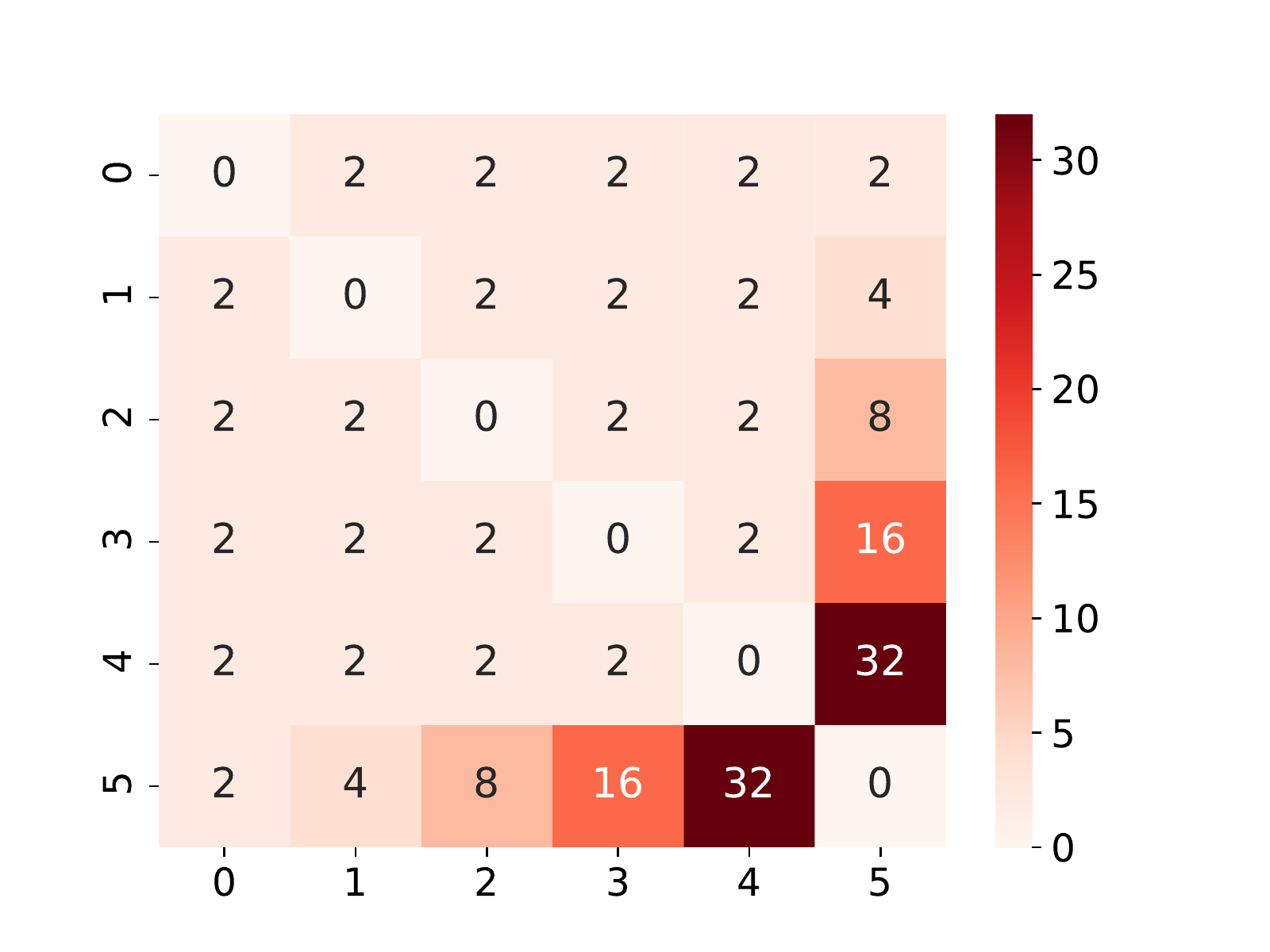}
(a)
\end{minipage}%
\begin{minipage}[t]{0.3\textwidth}
\centering
\includegraphics[width=\textwidth]{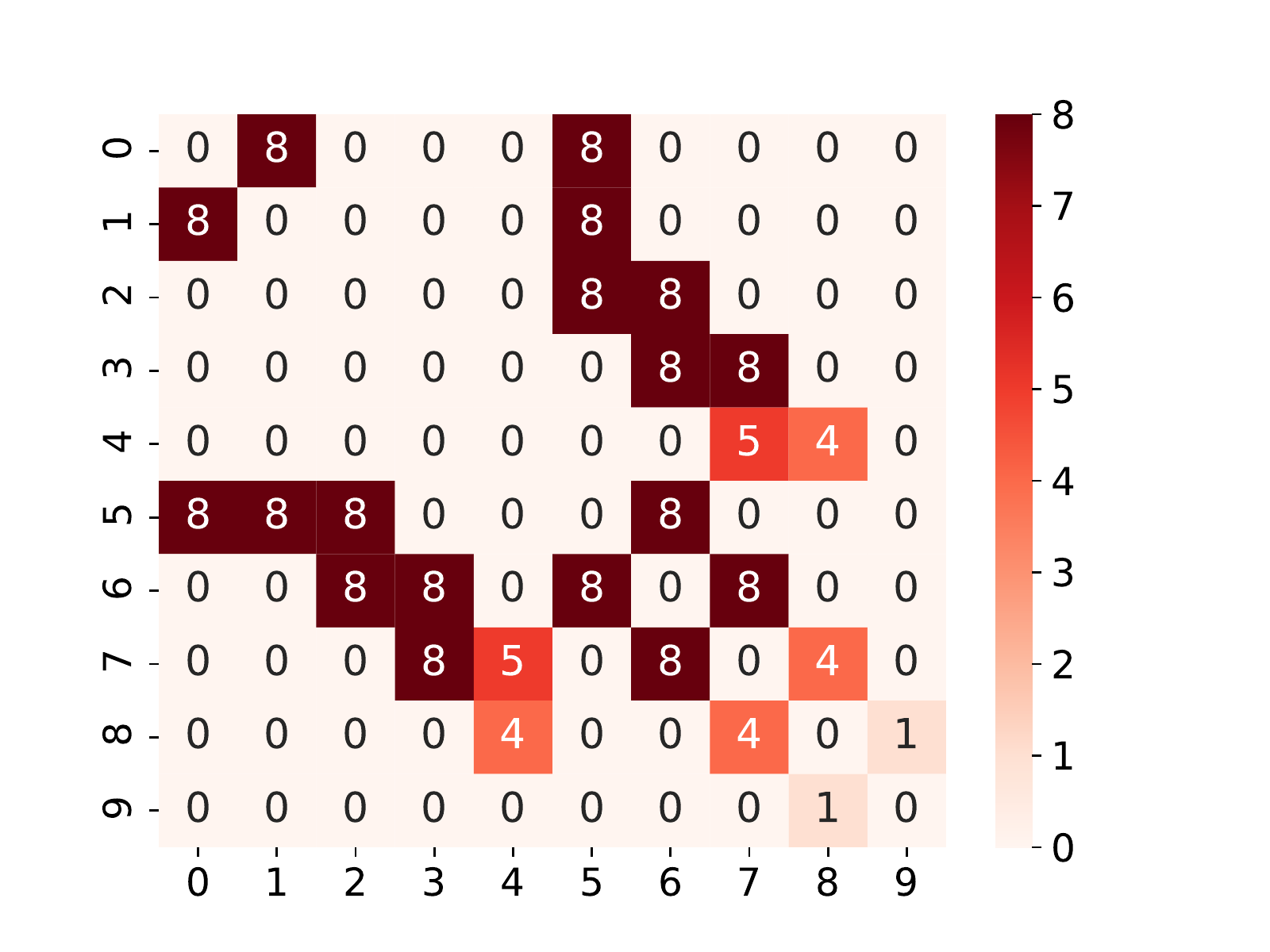}
(b)
\end{minipage}
\begin{minipage}[t]{0.3\textwidth}
\centering
\includegraphics[width=\textwidth]{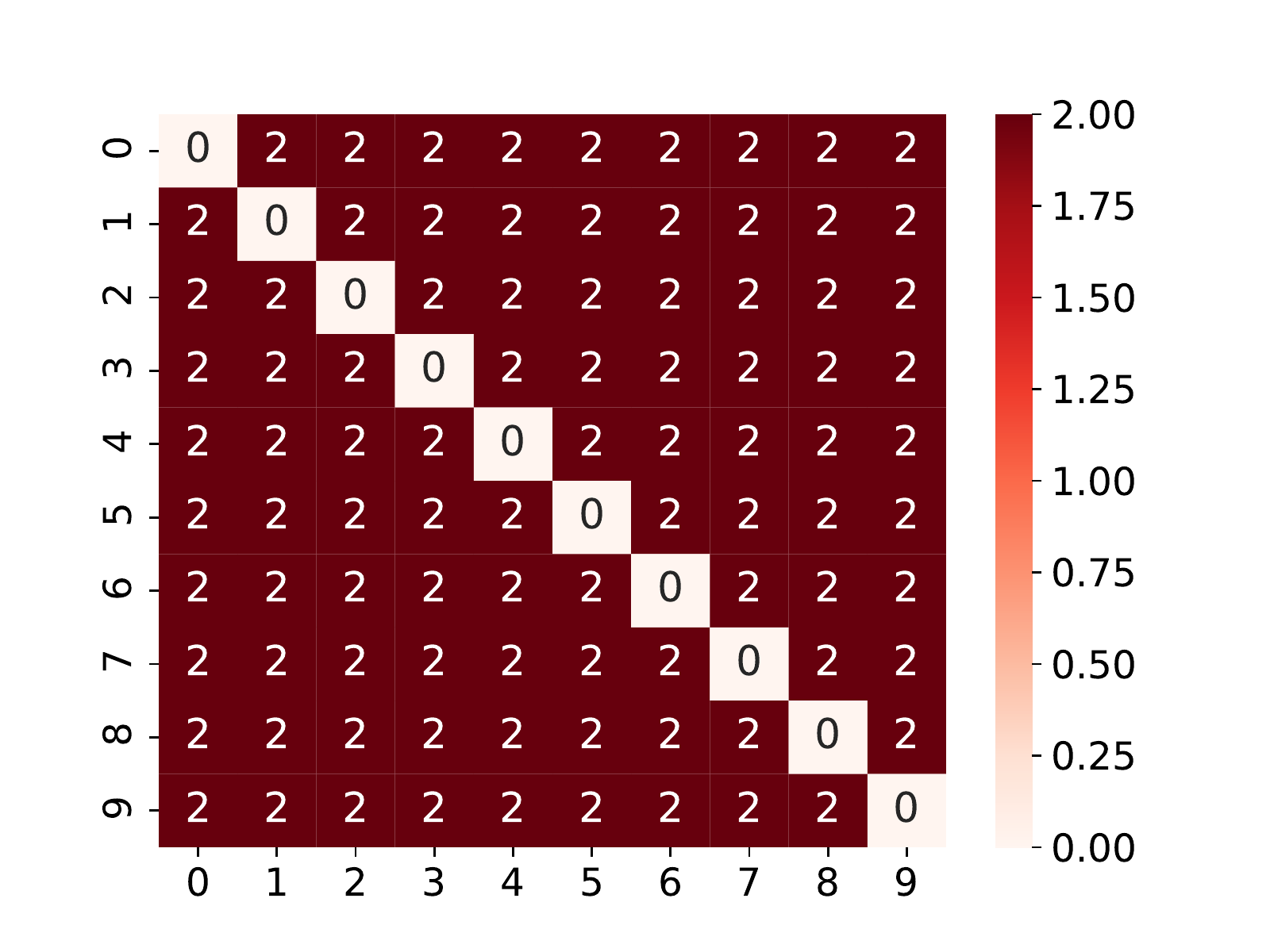}
(c)
\end{minipage}
\caption{\begin{footnotesize}\textbf{The adjacency matrix for phase estimation, Grover's search algorithm and Quantum Fourier Transformation(QFT).} 
(a) It can be seen that almost all of the high connection weight is concentrated on vertex $q_4$. 
(b) The connection is on the off-diagonal, the adjacency matrix forms a chain-like structure.
(c) All edges have the same weight.\end{footnotesize}}\label{fig different CCG}
\end{figure*}

\textbf{Maximum degree} 
The essence of the two-qubit gate is to entangle the control and target qubits \cite{chow2007,wang2016,liu2006,monz2011}. 
In superconducting system, the physical coupler that create entanglement will interfere the working frequency of the qubit
\cite{AQuantumEngineer'sGuidetoSuperconductingQubits,wallraff2004,chiorescu2004, brink2018,krinner2020,magesan2018}. 
Thus fabricating entangled structure is non trival.
There is a fully entangled structure \cite{song2017} and the largest number of entangled qubits so far is 10. But its drawback is that the entanglement cannot be closed,
which will create difficulties in the control of larger-scale quantum processor. Therefore, the current architecture still tends to establish entanglement between two qubits. 
Limited by the state of the art, the number of couplers on one qubit cann't be too large, i.e., there is a maximum degree of vertex in PCG.
In this paper, the maximum degree is set to be 6 because the largest degree of 2d lattices \cite{simon} and the achievable highest degree of structures in \cite{li2020} are 6.

\textbf{Planarization} 
Fabricating the 2d superconducting quantum processor is a mature technology \cite{boixo2018,arute2019}.
Thus all qubits and couplers are placed on one layer and the couplers cannot cross each other in space.
In the future, when multi-layer processor is available, some qubits or couplers can be placed on another layer.
In this situation, for the couplers on the same layer, circuit trnasformation is still of great essential 
\cite{chang2019,he2020} and SPQPD can play a significant role.
For this reason, a legitimate PCG should have a way of embedding to 2d plane without edge-interaction except at vertexes,
i.e., the PCG should be a planar graph.
\end{small}
\subsection*{\begin{normalsize}Profiling the Coupling Information\end{normalsize}}

\begin{figure*}[htbp]
\centering
\begin{minipage}[t]{0.23\textwidth}
\centering
\includegraphics[width=\textwidth]{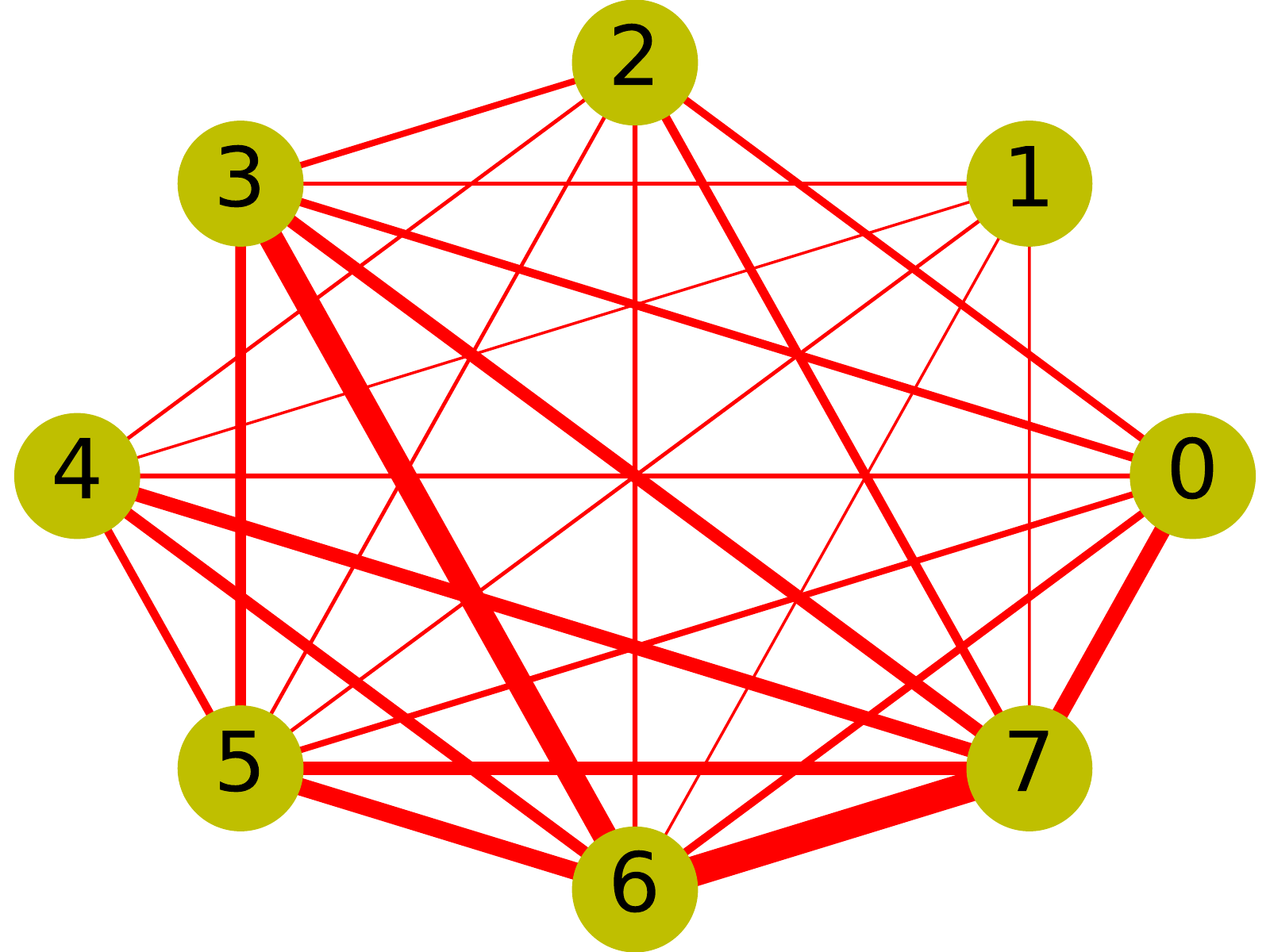}
(a)
\end{minipage}%
\begin{minipage}[t]{0.23\textwidth}
\centering
\includegraphics[width=\textwidth]{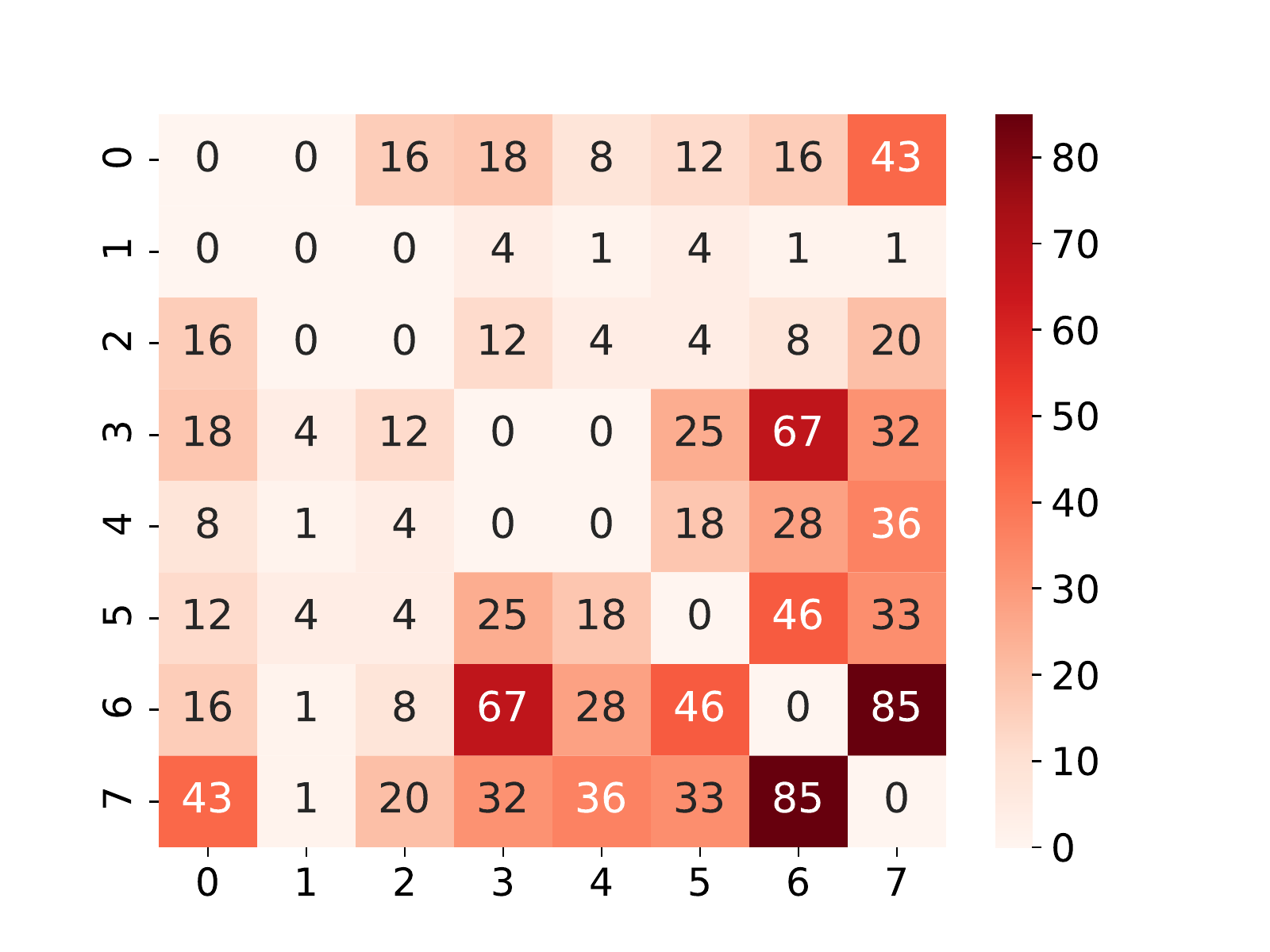}
(b)
\end{minipage}
\begin{minipage}[t]{0.23\textwidth}
\centering
\includegraphics[width=\textwidth]{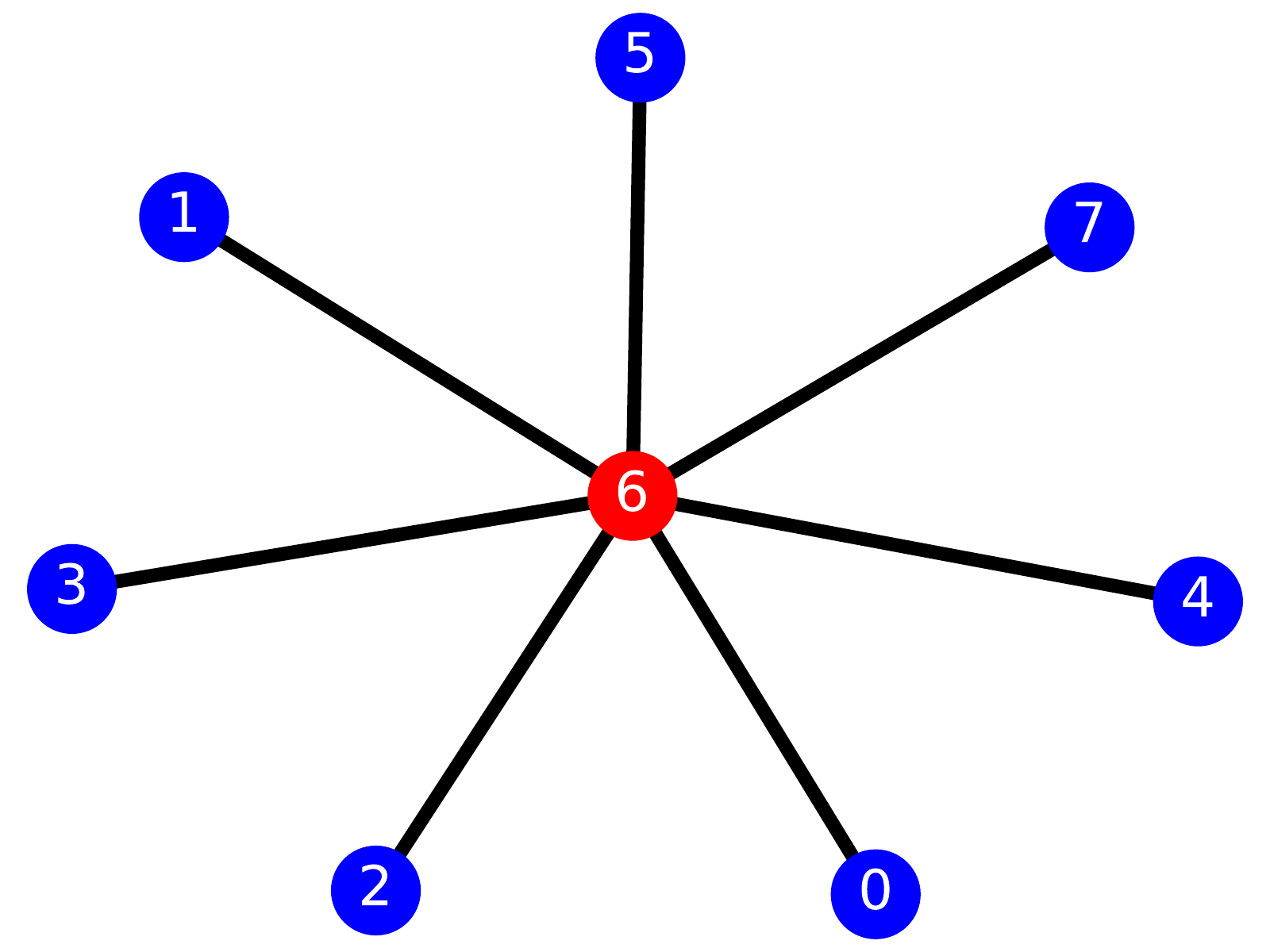}
(c)
\end{minipage}
\begin{minipage}[t]{0.23\textwidth}
\centering
\includegraphics[width=\textwidth]{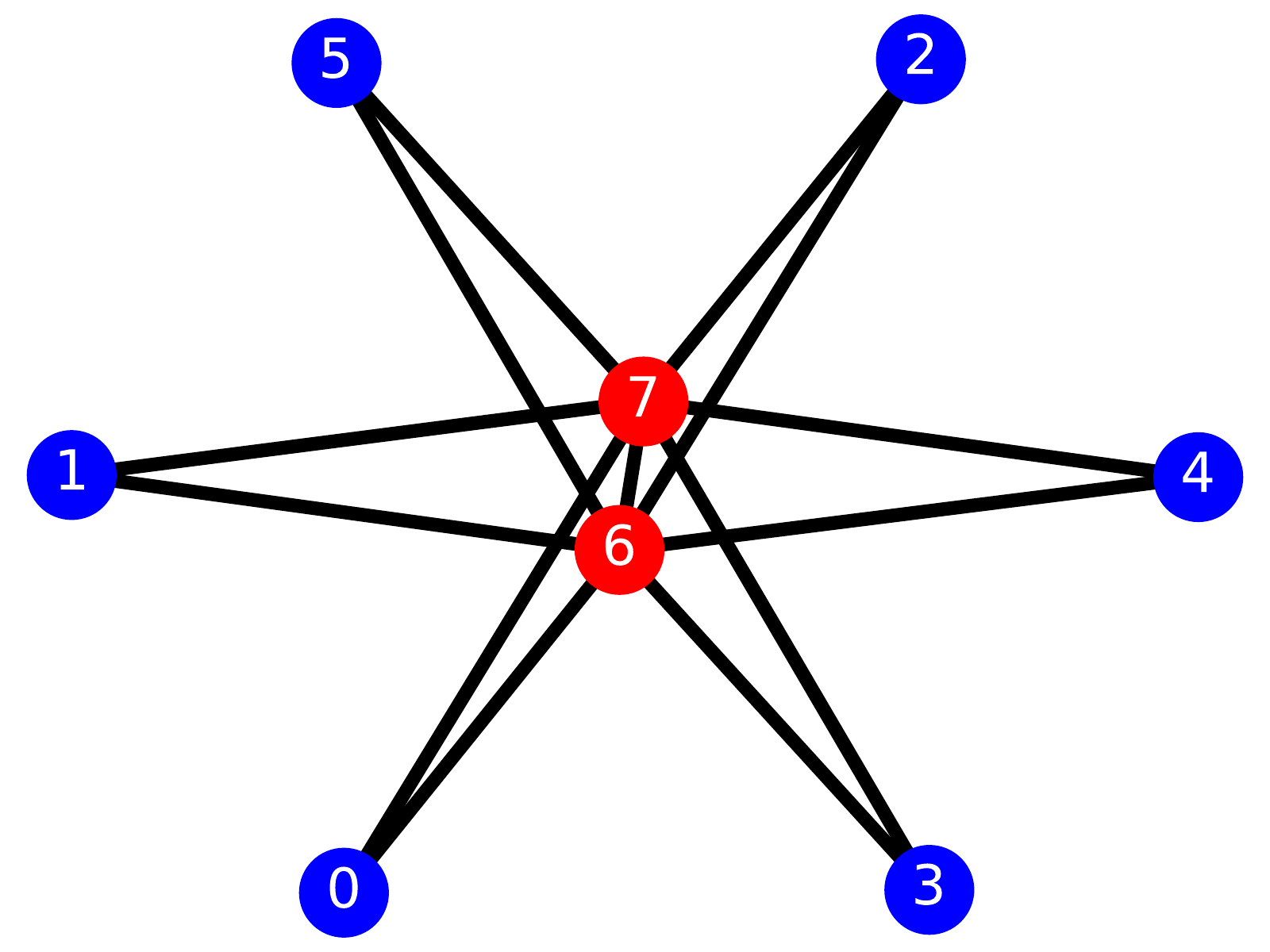}
(d)
\end{minipage}
\begin{minipage}[t]{0.23\textwidth}
\centering
\includegraphics[width=\textwidth]{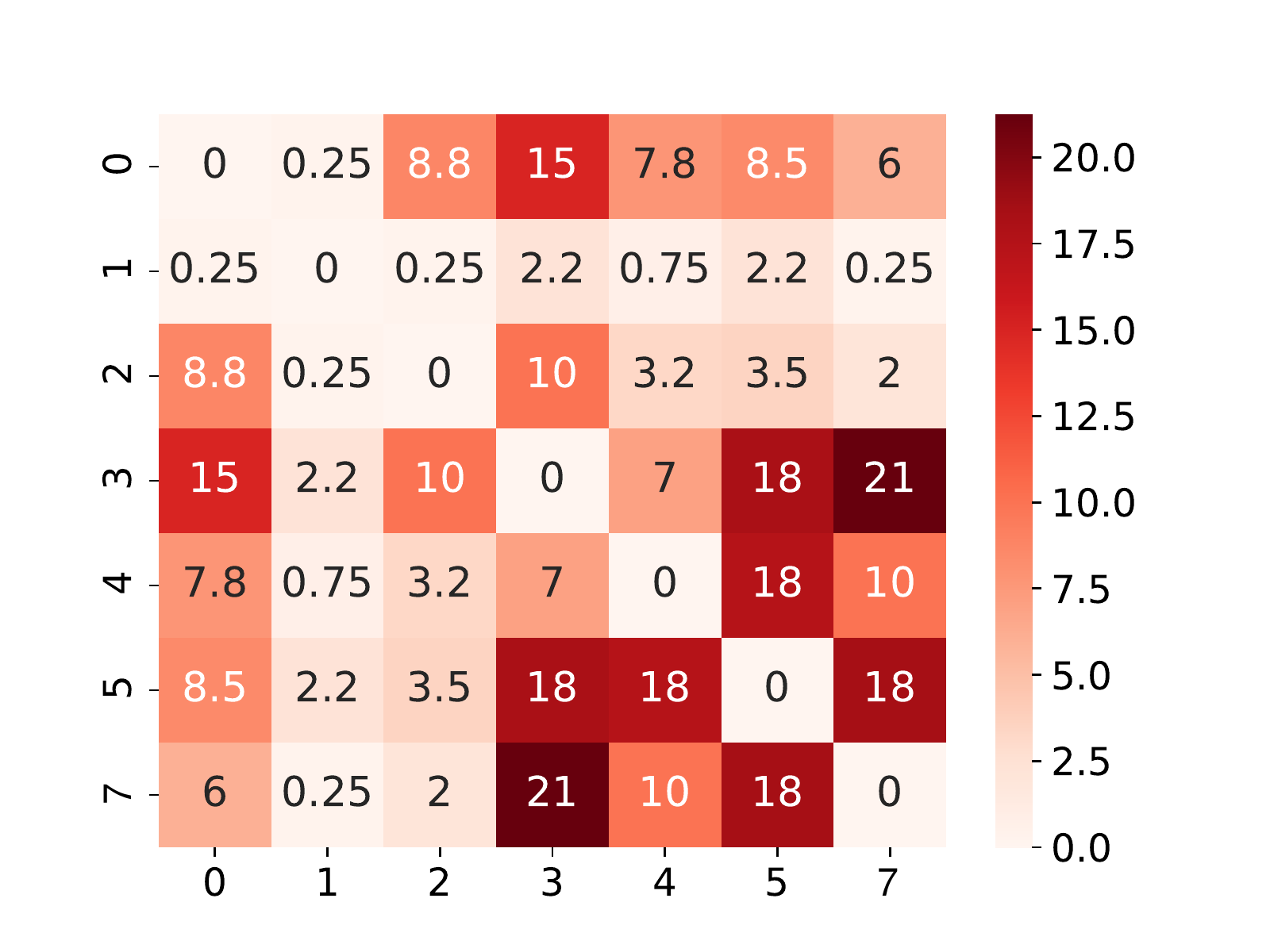}
(e)
\end{minipage}
\begin{minipage}[t]{0.23\textwidth}
\centering
\includegraphics[width=\textwidth]{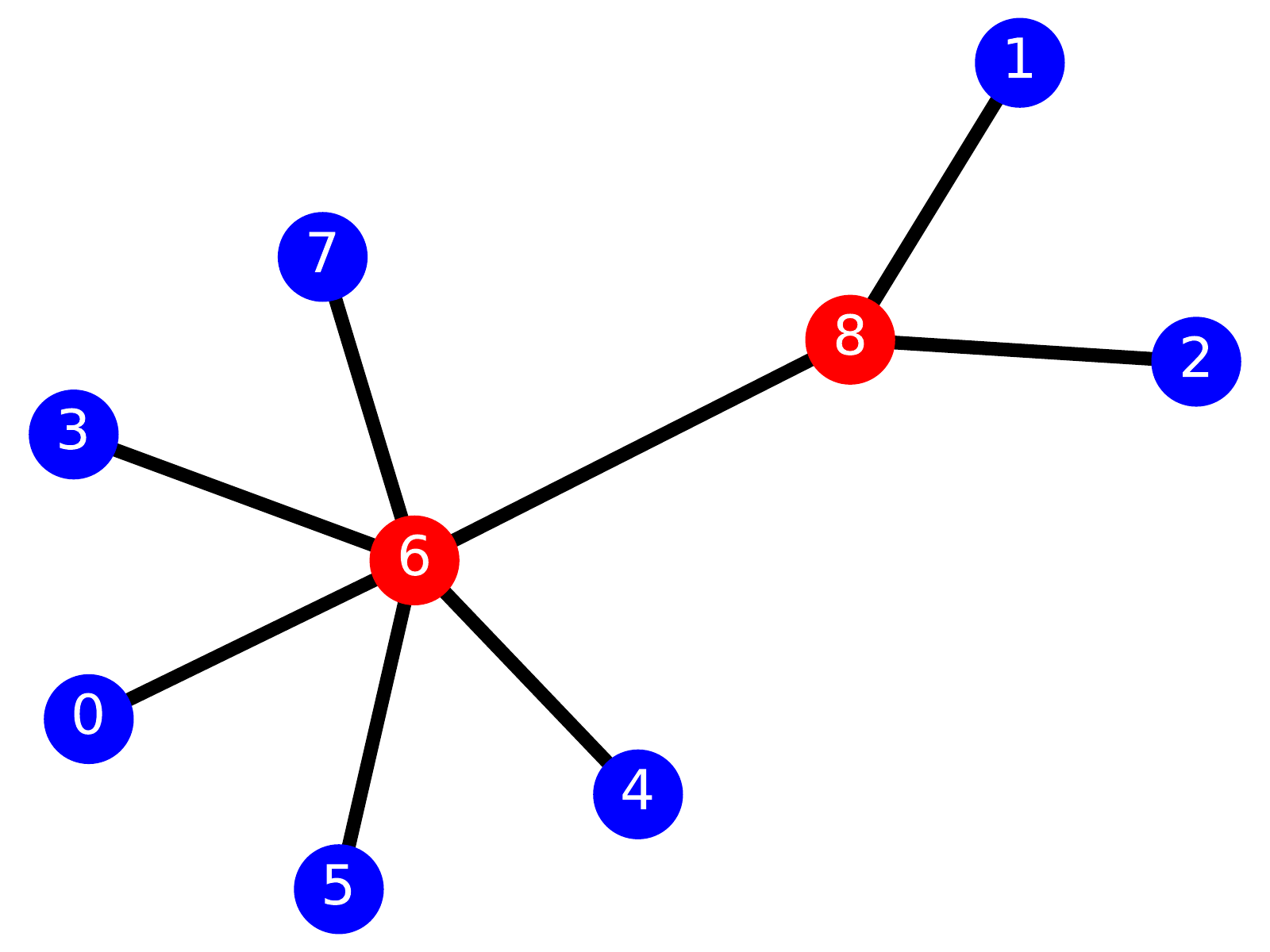}
(f)
\end{minipage}
\begin{minipage}[t]{0.23\textwidth}
\centering
\includegraphics[width=\textwidth]{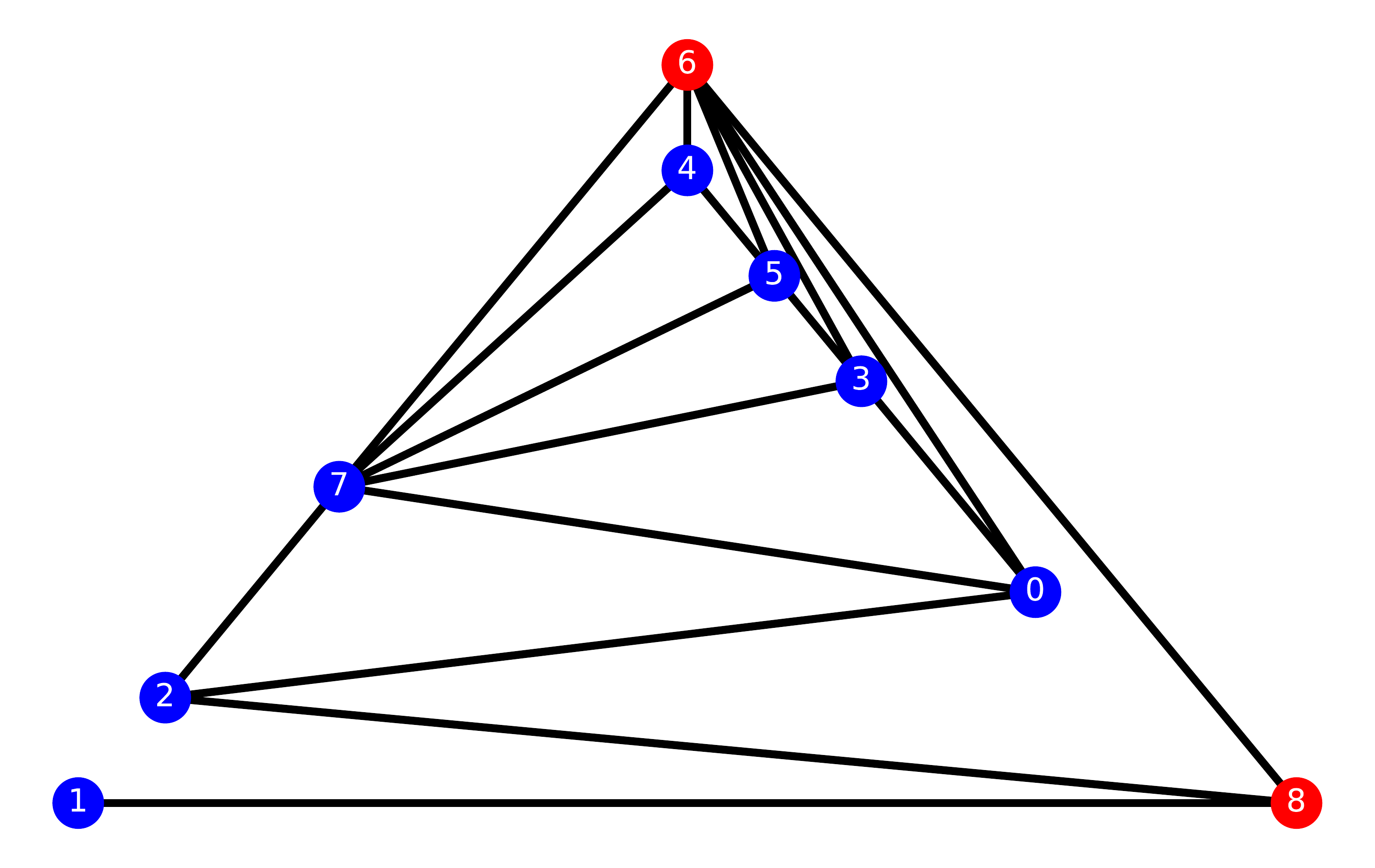}
(g)
\end{minipage}
\caption{\begin{footnotesize}\textbf{Example of the whole PCG design.} (a) The original CCG. (b) The corresponding adjacency matrix of (a). (c) Prunning result of $N=1$.
(d) Prunning result of $N=2$. (e) Interaction matrix $I$. (f) Result of splitting the media vertex and weight allocation. (g) The final CCG
which is also a legitimate PCG.\end{footnotesize}}\label{fig results}
\end{figure*}

\begin{small}
To design the quantum processor, we should profile information about the quantum circuit.
The number of two-qubit gates between vertexes pair is the key information to bridge the quantum circuit and processor, 
because there are a large number of two-qubit gates in a circuit while NISQ processor has extremely limited coupler resources. 
Designing a quantum processor according to the number of two-qubit gates is expected to dramatically reduce the number of extra gates. 
For this reason, two-qubit gates information should be profiled.

Fig.\;\ref{fig different CCG} are three examples of profiling the coupling information of a quantum circuit. 
These examples indicate the existence of different CCG patterns.
Fig.\;\ref{fig different CCG}(a) is an example of the phase estimation algorithm. Most of the weight is concentrated between $q_4$ and other vertexes. 
Fig.\;\ref{fig different CCG}(b) is an example of Grover's search algorithm.
The matrix elements are concentrated on the off-diagonal, thus the corresponding CCG forms a chain-like structure. 
Fig.\;\ref{fig different CCG}(c) is an example of the Quantum Fourier Transformation(QFT) algorithm. All edges are equally weighted.

Fig.\;\ref{fig different CCG} illustrates that the quantum processors can be customized for different circuits. E.g., the $q_4$ in \cref{fig different CCG}(a) need more coupler
resources, PCG of \cref{fig different CCG}(b) should be designed as a chain structure and the coupler resources in \cref{fig different CCG}(c) should be arranged more uniformly.
\end{small}
\subsection*{\begin{normalsize}Prunning\end{normalsize}}
\begin{small}
The second step is prunning the edges.
After profiling, we get a CCG. 
Since the CCG often violates the constraints, fabricating quantum processor with the original CCG's structure is intractable. 
Pruning some edges will reduce the degree, eliminate the intersections and make the CCG meet the constraints.
To answer the question that which edges need to be pruned, we present the following facts.  

First, if the edge is pruned, the two-qubit gates between the corresponding vertexes cannot execute directly, and swap gates are required.
Second, since the weight is the number of two-qubit gates, pruning the edges with smaller weight can reduce the number of extra swap gates.
Third, the vertex connected to more edges with larger weight should have more coupler resources.
Finally, the pruned graph must be planarizable.

Hence in this section, our goal is to modify the CCG into a planar graph with only a few important vertexes 
with high degree. It is sparser than the original CCG, more flexible and maybe more similar to the original CCG than the PCG with lattice-graph structures. 
And a PCG that meets the constraints and is as similar to the original CCG as possible can be designed in the further step.
\end{small}
\subsubsection*{\begin{small}Ranking Vertexes\end{small}}\label{subsect noderank}
\begin{small}
Here, we rank the vertexes according to their importance.
The key idea of our method is giving more coupler resources to the more important vertex. Because the connection information(detail discussions
are in the ``METHODS" section) such as degree and weight of the connected edges of each vertex are different, 
it is natural to infer that the importance of vertexes is different. We define a metric of vertex importance in the ``METHODS" section.
Under the metric, if the edges connected to the less important vertex were pruned, fewer swap gates will be inserted in the transformation process.

Fig.\;\ref{fig results} is an example design process from original CCG to the final legitimate PCG.
This CCG is generated from a circuit provided by \cite{Gadi2016}. The sum of the edge weight connected to $q_6$ is greater than that of other's,
so it is an important vertex that needs more coupler resources. 
After the ranking subroutine, the ranking result is $q_6,q_7,q_5,q_3,q_0,q_4,q_2,q_1$. 
\end{small}

\subsubsection*{\begin{small}Prunning Based on the Media Vertexes\end{small}}\label{sect pruning}
\begin{small}
Now we should complete the prunning process according to the ranking result. 
Based on the ideas proposed above, we shouldn't change the vertexes with high rankings, but prune the edges of the vertexes
with lower rankings. 

The original CCG \cref{fig results}(a) violates the constraint of planarization, thus the pruning subroutine is needed. 
The ranking result shows that the most important and the second important vertex are $q_6$ and $q_7$. 
Fig.\;\ref{fig results}(c) is the pruning result of the media vertex number $N=1$, where $q_6$ is chosen as media vertex.
Fig.\;\ref{fig results}(d) is the result of $N=2$, where $q_6$ and $q_7$ are both chosen as media vertexes. These media vertexes
are not only the most important, but also serve as the path to execute the two-qubit gate between non adjacent vertexes. 
After choosing the media vertexes, the edge that connect $q_i,q_j$ is pruned if
$\{q_i,q_j\}\cap media\_vertex\_set=\emptyset$, where $media\_vertex\_set$ is the set of media vertexes.
The pruned edges are stored in a \emph{\textbf{recover\_set}} that is useful in the ``Recovering" section. 
The question of how many media vertexes should be selected is discussed in the ``METHODS" section.
\end{small}
\subsection*{\begin{normalsize}Handling the High Degree Vertex\end{normalsize}}\label{sect handlehigh}
\subsubsection*{\begin{small}Splitting the High Degree Vertex\end{small}}\label{sect split}
\begin{figure}[htbp]
\centering
\begin{minipage}[t]{0.4\textwidth}
\centering
\includegraphics[width=\textwidth]{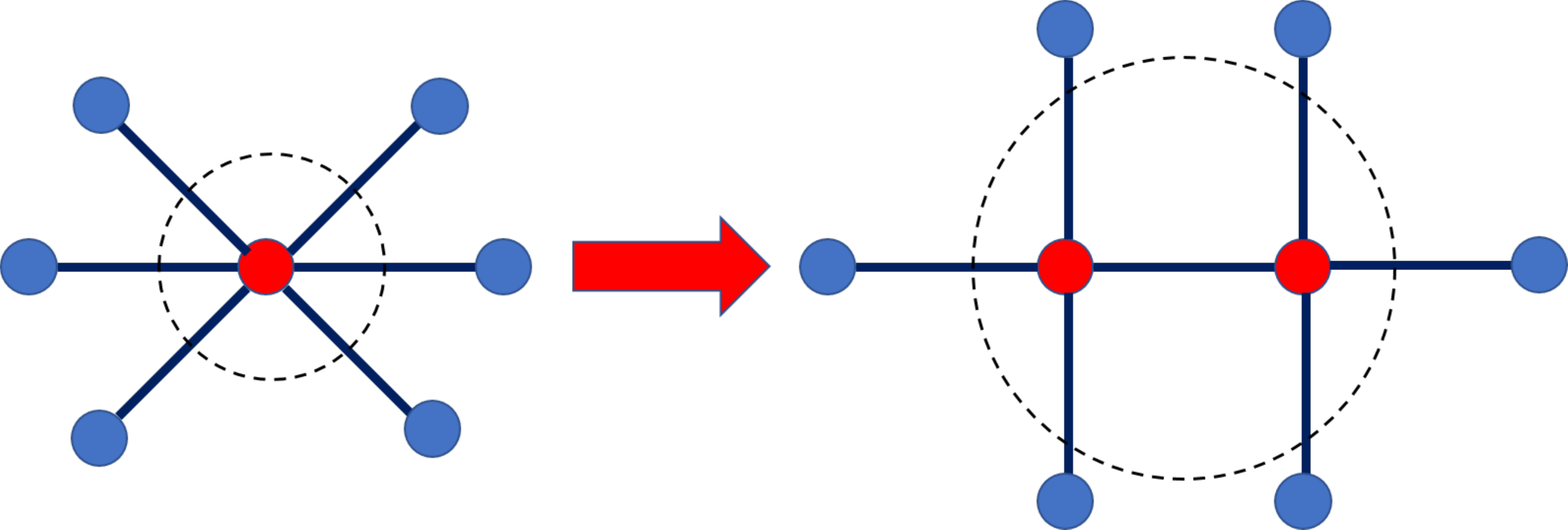}
(a)
\end{minipage}
\centering
\begin{minipage}[t]{0.4\textwidth}
\centering
\includegraphics[width=\textwidth]{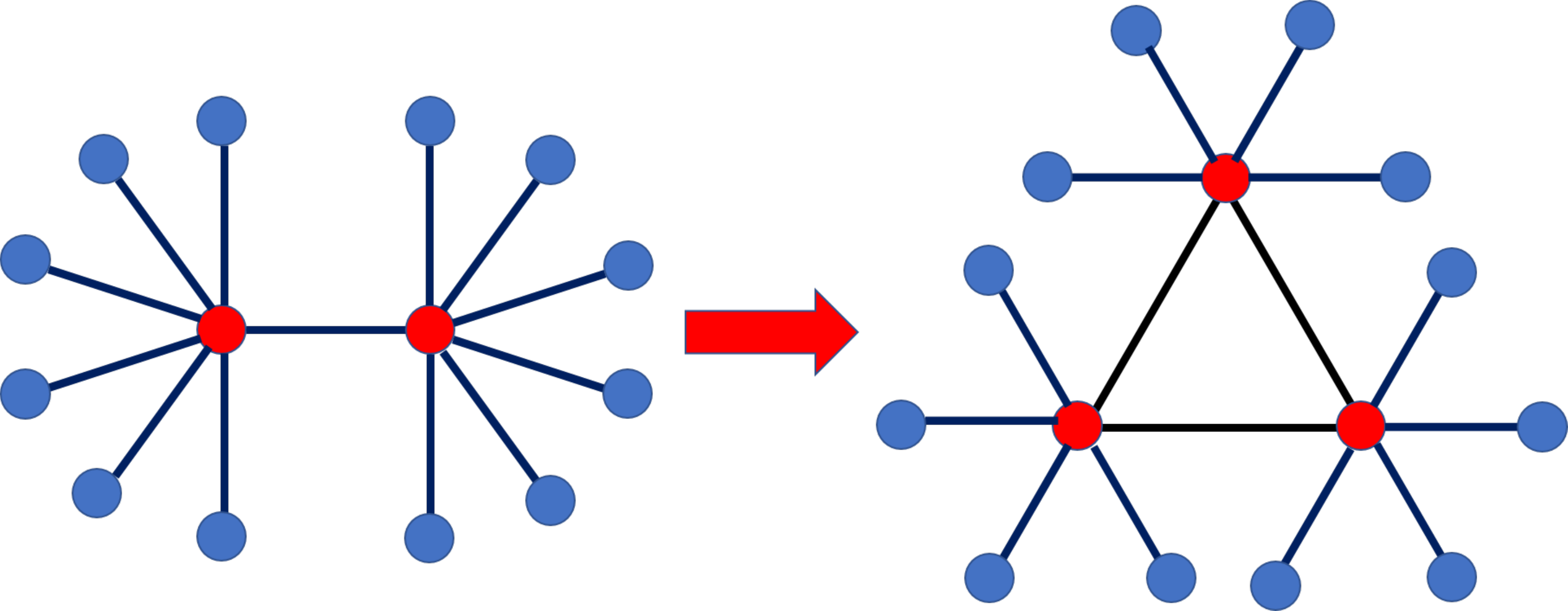}
(b)
\end{minipage}
\caption{\begin{footnotesize}\textbf{Add ancilla qubits and change the media structure into a structure with multiple vertexes.} (a) The degree is reduced from 6 to 4.
(b) The degree is reduced from 7 to 4.\end{footnotesize}}\label{fig changeconnect}
\end{figure}

\begin{small}
The third step is handling the vertexes whose degree violate the maximum degree constraint.
After pruning, although we have eliminated the intersections in the CCG and pruned some edges that violate degree constraint, the degrees of the media vertexes in the graph may 
still violate the constraint of the maximum degree.
For this reason, the third step of SPQPD is adding ancilla vertexes and changing the media vertex $v$ into a graph called media structure
composed of multiple vertexes $\{v_1,v_2,\cdots,v_n\}$. The logical qubit contained by the media vertex $v$ now corresponds to one vertex $v_i$ in media structure at a time.
We connect a part of edges of media vertex to the ancilla vertexes to reduce the excessive degree.
Fig.\;\ref{fig changeconnect}(a) and \cref{fig changeconnect}(b) are two examples of reducing degree by adding ancilla qubits. 
\end{small}

\begin{small}
This step is defined as splitting media vertex. The media structure must satisfy the following four conditions.
First, the media structure should be a connected graph. Therefore, for a media structure with $n$ vertexes, the number of edges $e$ of the media structure should satisfy
$e\geqslant n-1$.
Second, as in \cref{fig changeconnect}(a), from the external perspective of the media structure, media structure is a black box that behaves like the media vertex.
Therefore, media structure should contain sufficient free connections for the non-media vertexes connected by the media vertex, i.e., $k\leqslant Dn-e,\label{eq media cond2}$
where $D$ is the maximum degree, $k$ is the number of non-media vertexes connected by original media vertex.
Third, the graph of media structure should meet the technical constraints for manufacturable PCGs.
Forth, the number of vertexes $n$ should be as few as possible because increasing the number of ancilla qubits will increase the difficulty of processor manufacturing.
The subroutine which can search for media structures based on these conditions will be discussed in the ``METHODS" section.
\end{small}
\subsubsection*{\begin{small}Allocating the Non-Media Vertexes\end{small}}\label{subsect splitselect}

\begin{figure*}[htbp]
\centering
\begin{minipage}[t]{0.24\textwidth}
\centering
\includegraphics[width=\textwidth]{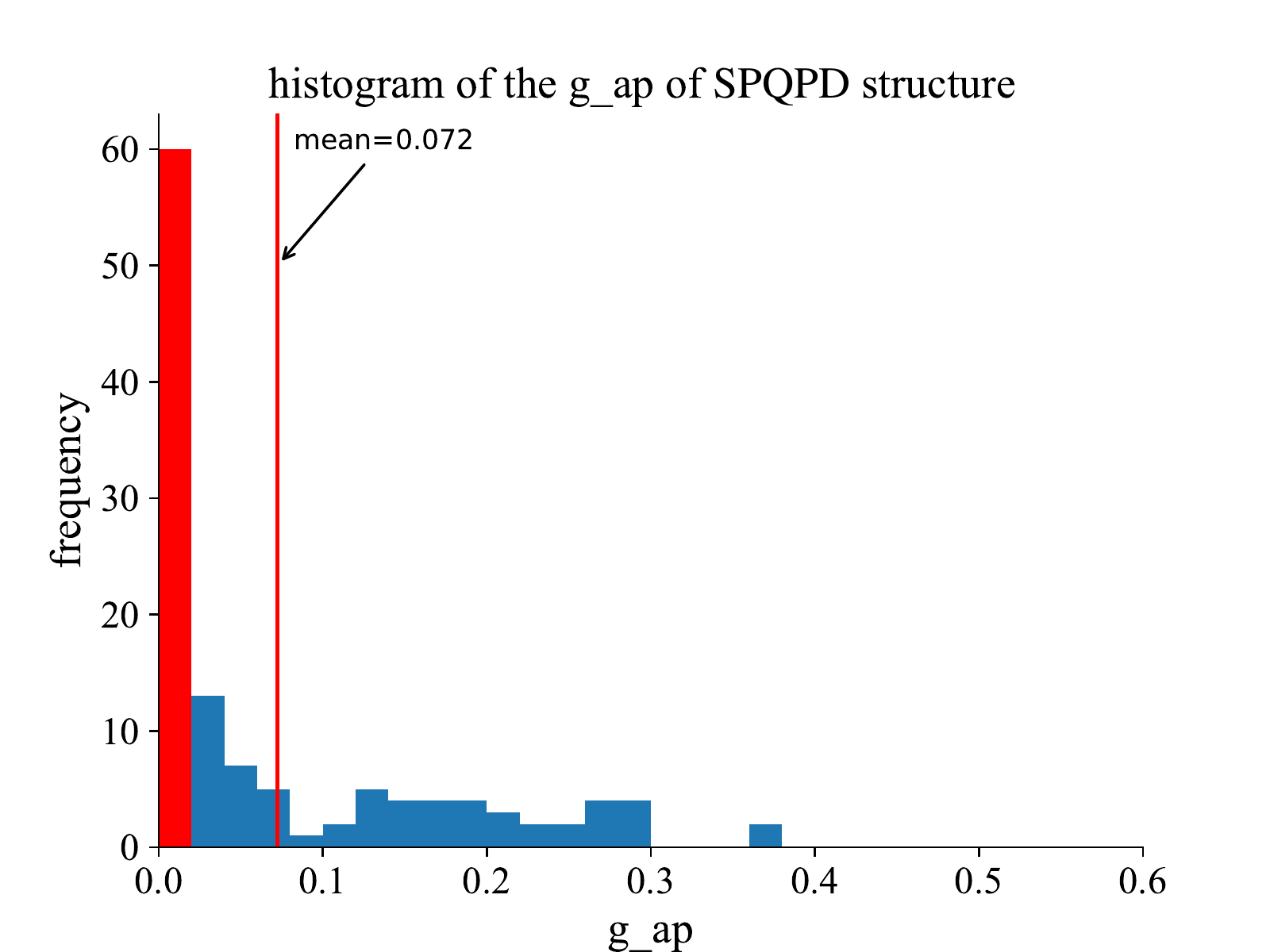}
(a)
\end{minipage}
\begin{minipage}[t]{0.24\textwidth}
\centering
\includegraphics[width=\textwidth]{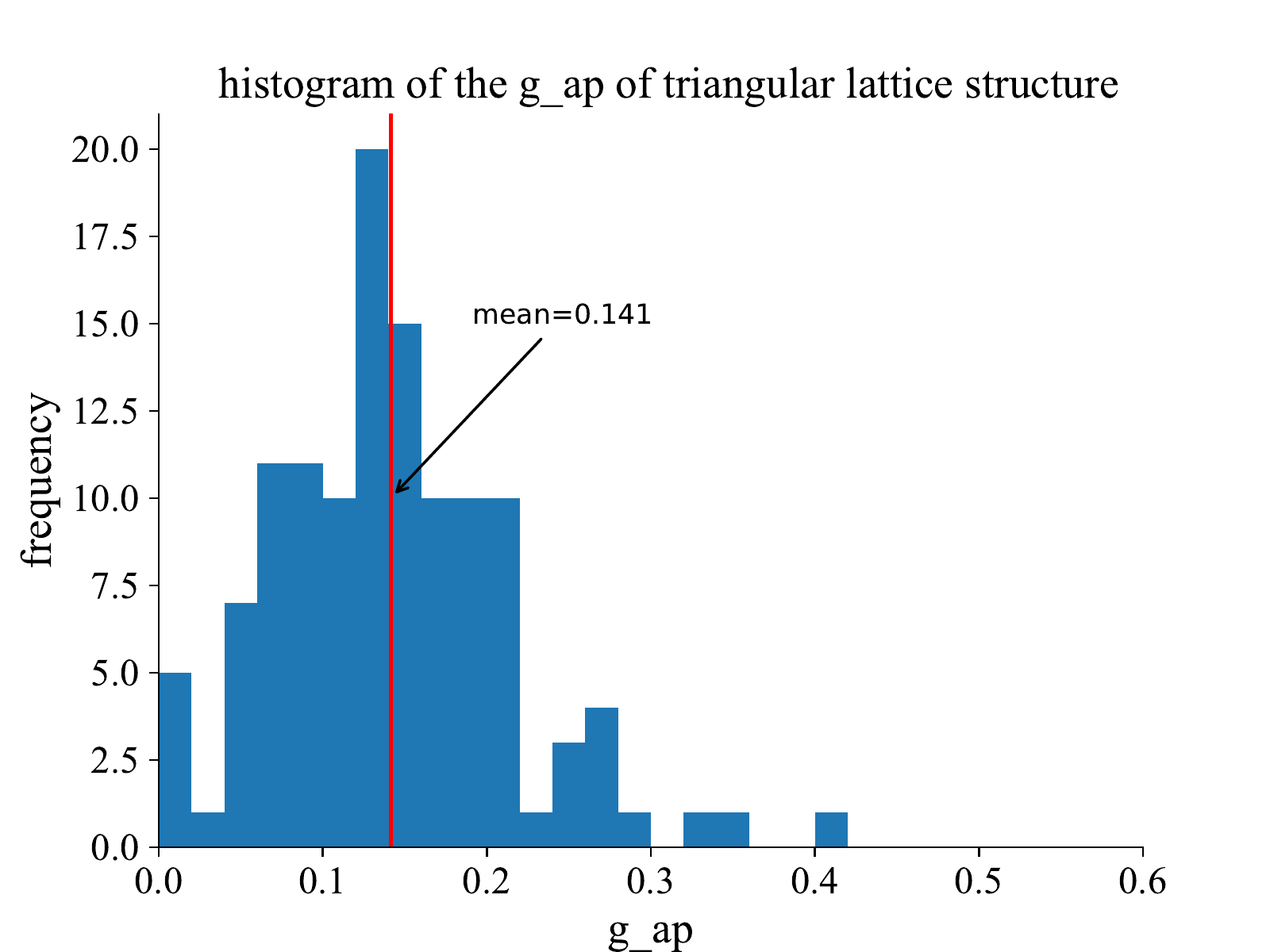}
(b)
\end{minipage}
\centering
\begin{minipage}[t]{0.24\textwidth}
\centering
\includegraphics[width=\textwidth]{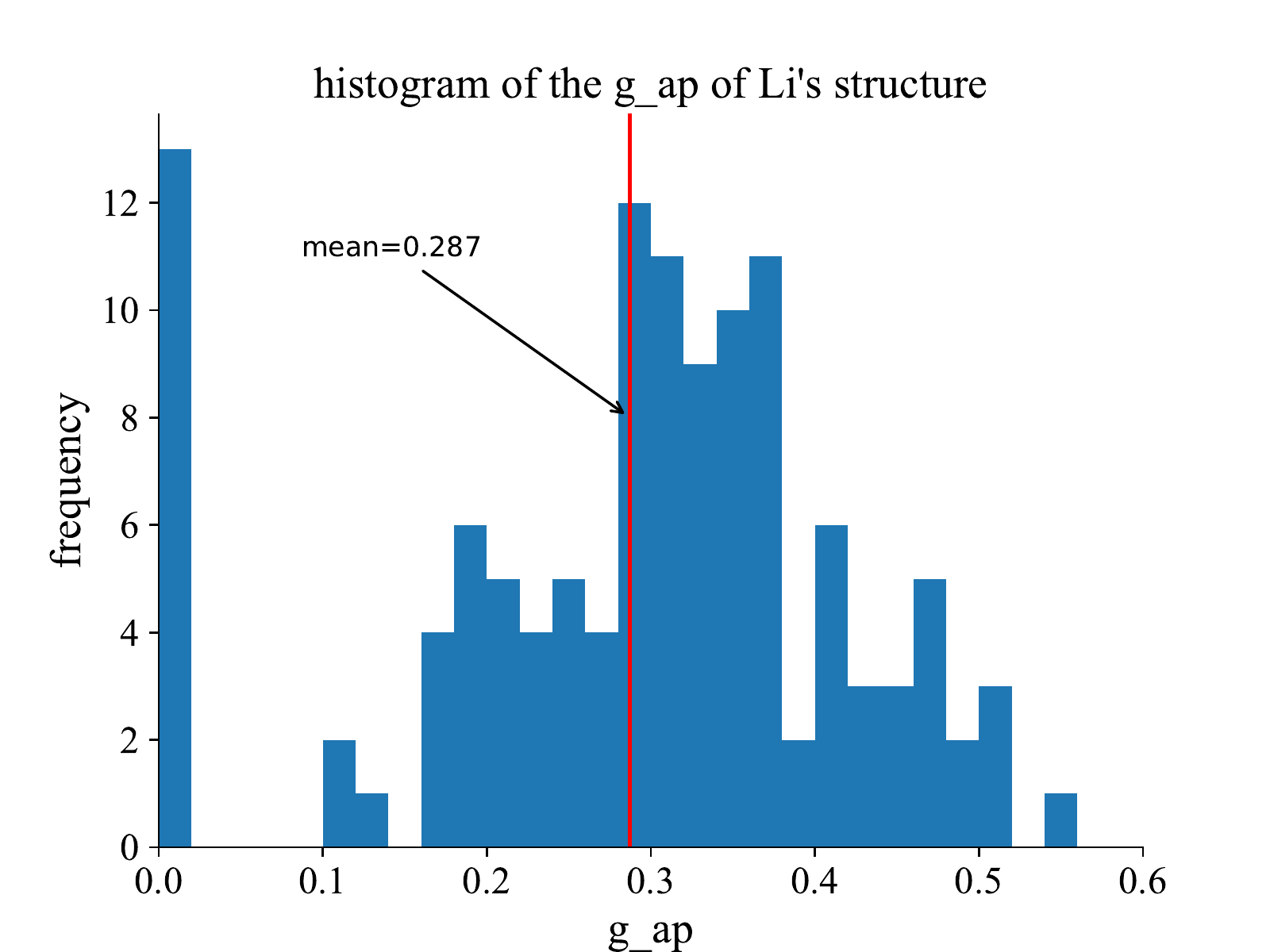}
(c)
\end{minipage}
\begin{minipage}[t]{0.24\textwidth}
\centering
\includegraphics[width=\textwidth]{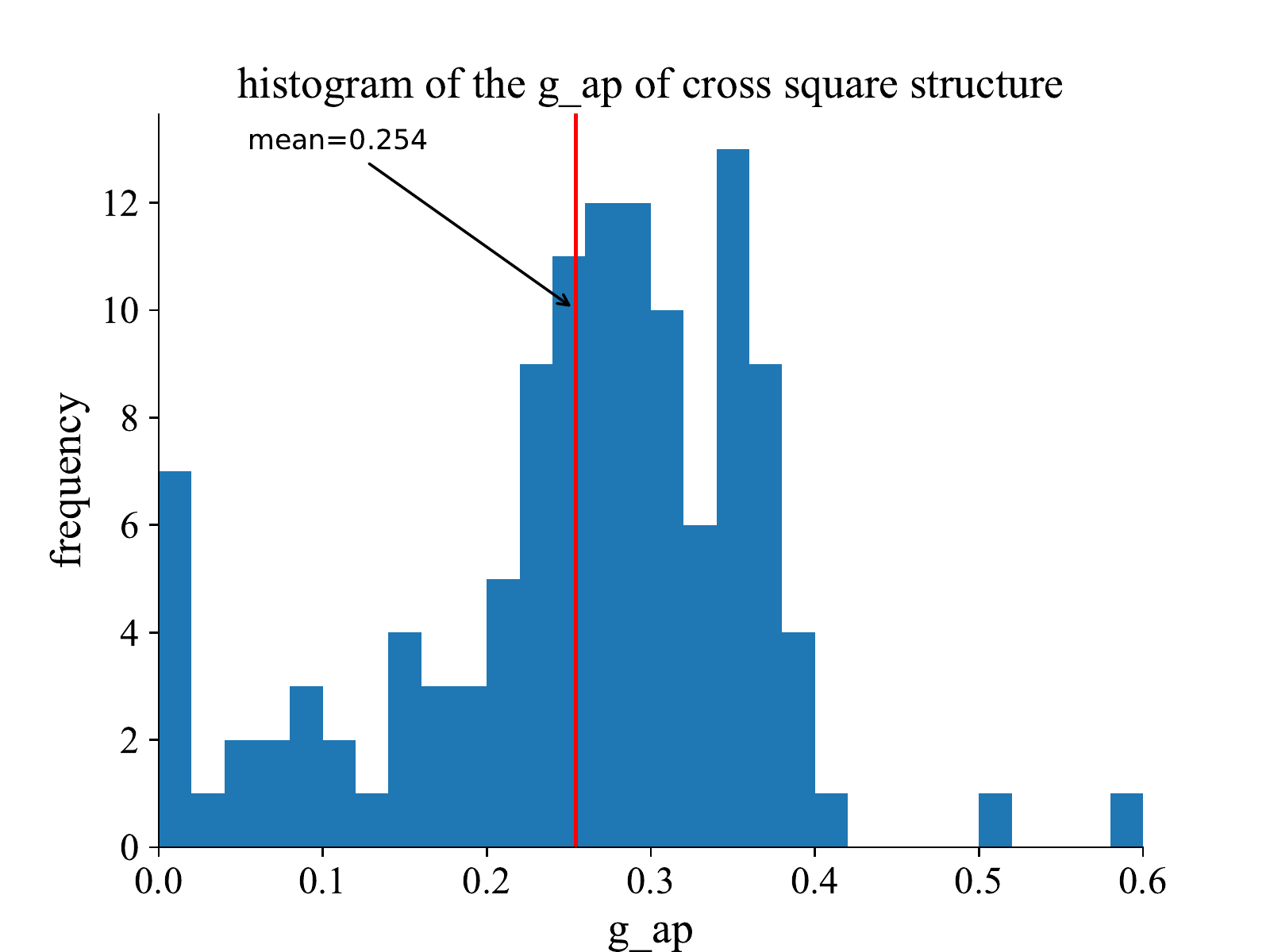}
(d)
\end{minipage}
\caption{\begin{footnotesize}\textbf{The histograms of $g_{ap}$.} 
The mean values $\overline{g_{ap}}$ are 0.072, 0.141, 0.287, 0.254. The distribution in (a) has the smallest mean value. 
The red bar in (a) has the largest frequency and it belongs to the smallest interval.
\end{footnotesize}}\label{fig performance}
\end{figure*}

\begin{figure*}[htbp]
\centering
\begin{minipage}[t]{0.4\textwidth}
\centering
\includegraphics[width=\textwidth]{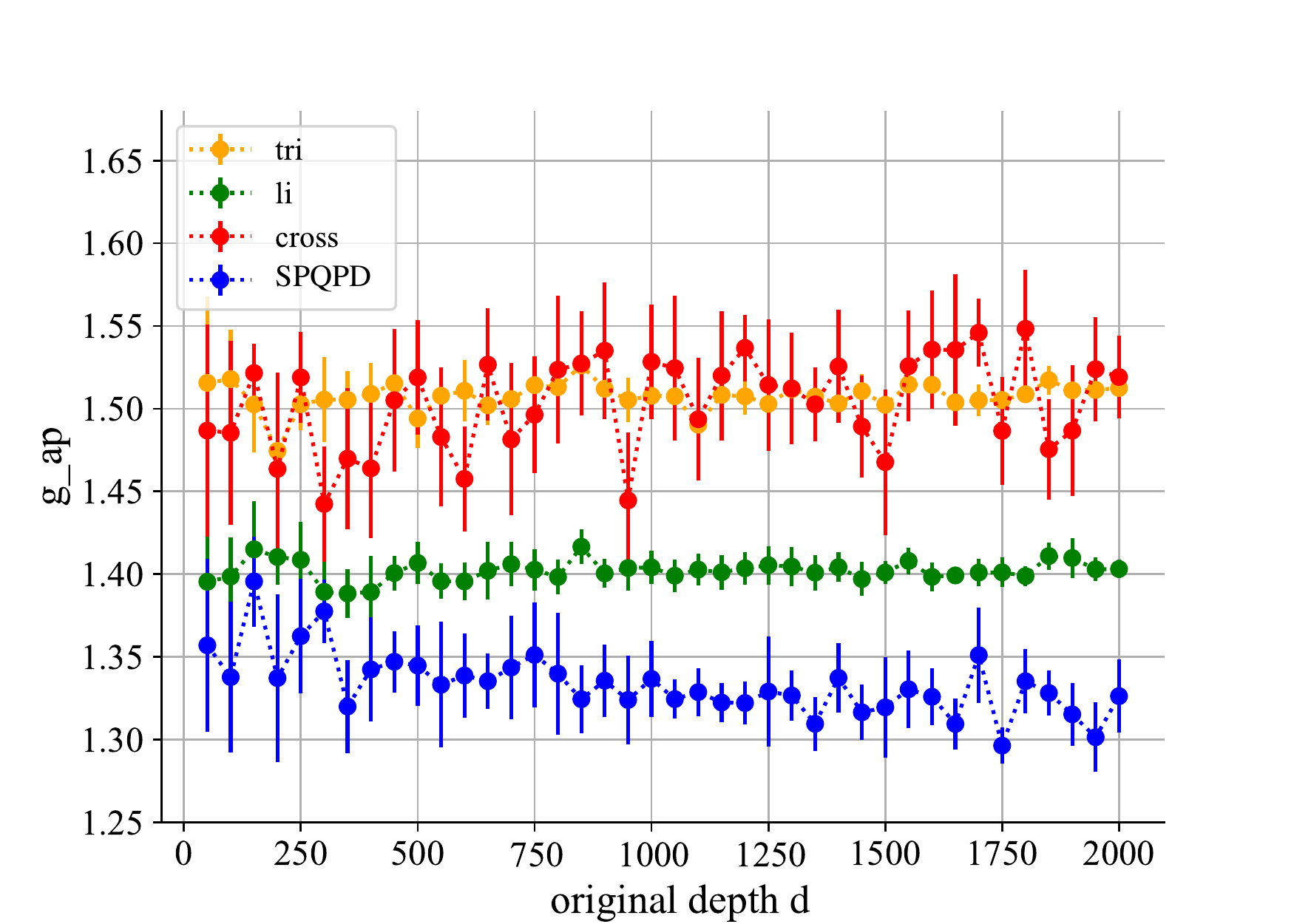}
(a)
\end{minipage}
\begin{minipage}[t]{0.4\textwidth}
\centering
\includegraphics[width=\textwidth]{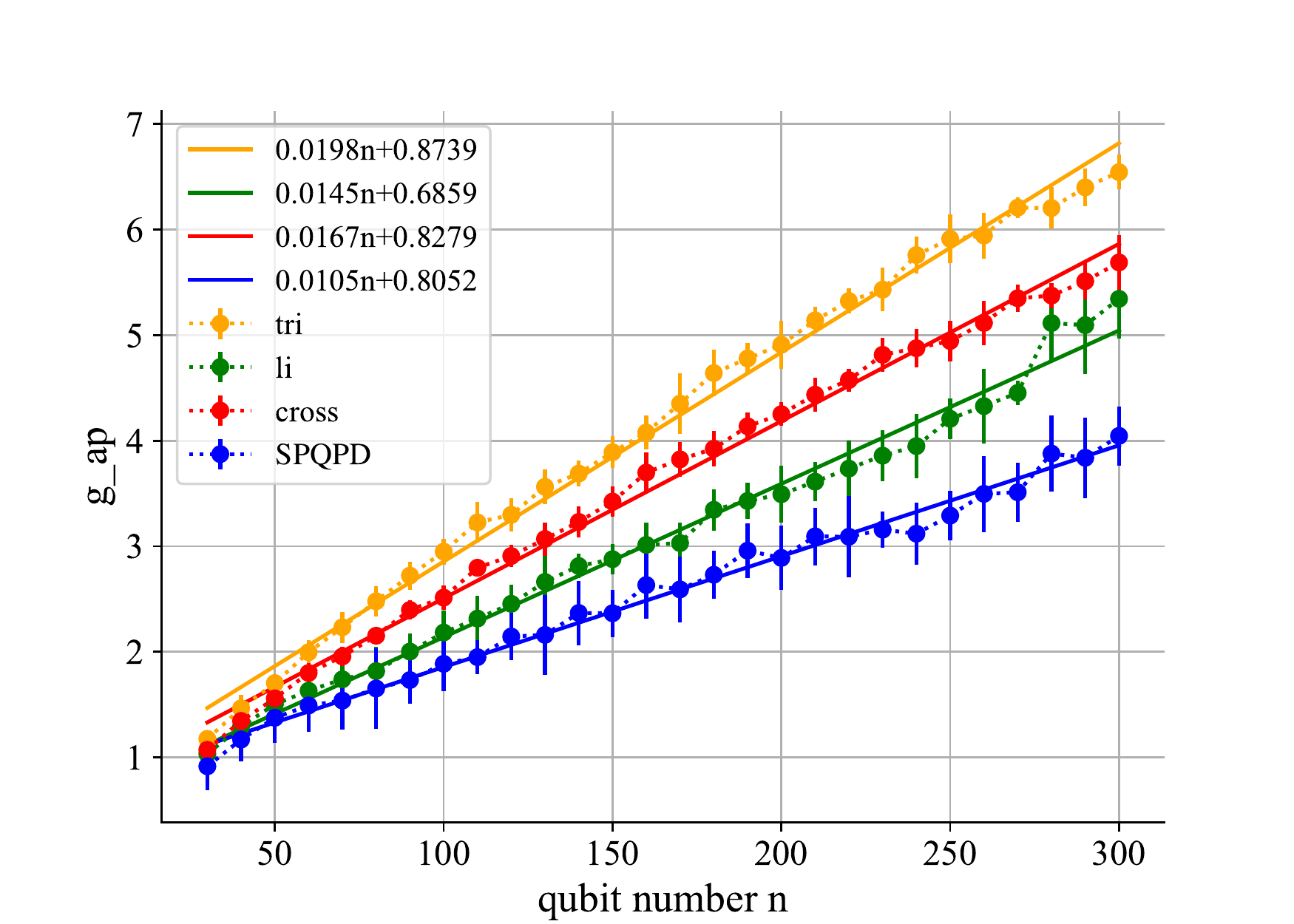}
(b)
\end{minipage}
\caption{\begin{footnotesize}\textbf{The relationships between $g_{ap}$ and original depth and qubit number.} 
(a) The relationship between $g_{ap}$ and original depth. SPQPD method has the smallest $g_{ap}$ within the depth range $[50, 2000]$. 
For SPQPD, the mean value of each error bar $\langle g_{ap}\rangle_d$ decreases as circuit depth gets larger. 
For the others, the corresponding character either increases or has no obvious trend. The qubit number is set to be 50.
(b) The relationship between $g_{ap}$ and qubit number. SPQPD method has the smallest $g_{ap}$ growth rate among all structures, where original depth is set to be 200.
\end{footnotesize}}\label{fig performance2}
\end{figure*}

\begin{small}
After replacing the media vertex by media structure, we need to reconnect the non-media vertexes to the vertexes in the media structure. This step is called weight allocation
and the allocation way is not unique. We should find an optimized allocation way to minimize the extra swap gates because of the following reasons.

First, because only one vertex $v_i$ in the media structure contains the logical qubit of the original media vertex $v$ at a time,
swap gates are required if $v_i$ doesn't directly connected to the non-media vertex $q_l$ 
when two-qubit gate $g$ between $q_l,v$ need to be executed immediately. 
To describe this, we define a matrix $S$ whose element is the sum of the number of the two-qubit gate blocks in two subcircuits only about vertexes $q_m,v$ and $q_l,v$.
Second, the non-media vertex are now connected by media vertex, swap gates are required
when two-qubit gate between non-media vertexes $q_m,q_l$ need to be executed. The farther they are on the graph, the more swap gates are required. 
The adjacency matrix $M$ can be used to describe this.

To describe how close $q_m,q_l$ should be allocated on the media structure,
we define an interaction matrix $I$ whose element is $I_{ml}=aM_{ml}+(1-a)S_{ml}$, where $a$ is the combinition coefficience.

Another matrix required is $C$ whose element is the shortest length between vertexes in media structure.
Our goal is to find an allocation map \emph{alloc} to let the score 
\end{small}
\begin{align}
&F=\sum_{ij,alloc[i]alloc[j]}C_{ij}I_{alloc[i]alloc[j]},\label{eq F}\\
&alloc:V\rightarrow 2^{\{q|q\in\text{neighbor of }v\}}
\end{align}

\begin{small}
\noindent get the minimum value, where $alloc$ is a map that allocates the neighbor of media vertex $v$ to the vertexes in the media structure.
$V$ is the set of vertexes in media structure and $v$ is the original media vertex.

In \cref{fig results}(c), we set $N=1$, the most important vertex is $q_6$. Thus after the pruning subroutine, $q_6$ is selected as media vertex.
$q_6$ violates the degree constraint and need to be replaced by a media structure. First, we try the media structure with two vertexes ${q_6,q_8}$ which meets the condition 
$1=e\leqslant Dn-k=6\times2-7=5$. Then, the non-media vertexes are allocated according to the matrix $I$. 
In \cref{fig results}(e), the elements in columns 3, 5, 7, 0 and 4 of $I$ is relatively larger than columns 1 and 2, thus allocating the former 5
vertexes to the same vertex $q_6$ will let the summation terms in \cref{eq F} with larger $I_{alloc[i]alloc[j]}$ become zero and reduce the value of $F$.
After $q_6$ reaching the degree constraint $D_{q_6}=6$, $q_1$ and $q_2$ have to be allocated to $q_8$,
because they have less interaction with the former vertexes, the impact that their allocation is far away from the former important vertexes is smaller.
Thus, the result is \cref{eq allo1}, \cref{eq allo2} and \cref{fig results}(f).
\end{small}
\begin{align}
&alloc[q_6]=\{q_5,q_0,q_3,q_7,q_4\},\label{eq allo1}\\
&alloc[q_8]=\{q_1,q_2\},\label{eq allo2}
\end{align}
\begin{small}
\noindent where $q_6,q_8$ are vertexes in media structure and the others are the non-media vertexes needed to be allocated.

The media vertexes with more than two vertexes have multiple non-isomorphic topological structures.
In this case, we should repeat the process above and find the structure with the smallest score $F$. 
\end{small}
\subsection*{\begin{normalsize}Recovering\end{normalsize}}\label{sect recover}
\begin{small}
The forth step recovering is to reconnect some pruned edges to the CCG.
In \cref{fig results}(f), vertexes except for $q_6$ are far from the maximum degree, this means that some edges in the \emph{recover\_set} can be recovered 
to the CCG. Based on the idea that changing the subgraph with smaller weight and degree has less impact, the edge with larger degree has higher recovery priority.

Fig.\;\ref{fig results}(g) is the final CCG of the routine. The structure of this CCG meets the constraints mentioned above, 
so it is also a legitimate PCG structure that can be implemented by the superconducting system.
\end{small}
\subsection*{\begin{normalsize}Comparing the Number of Extra Swap Gates\end{normalsize}}
\subsubsection*{\begin{small}Using 158 Benchmarks\end{small}}
\begin{small}
Finally, we use a series of circuits as benchmarks to test SPQPD method and compare it with other lattice-graph structures.
158 circuits are used for the experiment \cite{Gadi2016}. Triangular lattice structure, the 2d planar lattice structure that with the largest degree 6 \cite{simon},
structures designed by the method in \cite{li2020} and cross square lattice structure based on the same foundation as the Li's structure
are used as representatives of lattice-graph structures. 

We define the metric as the number of extra swap gates per input two-qubit gate $g_{ap}=\frac{g_{add}}{g_{ori}}$ 
where $g_{add}$ is the number of extra swap gates and $g_{ori}$ is the number of original two-qubit gate. 

Fig.\;\ref{fig performance}(a)-(d) are the histograms of $g_{ap}$ for four kinds of structures. The results show that on average, the SPQPD method has the best
performance and its average $\overline{g_{ap}}=0.072$. It is improved by $104.2\%$ compared with the second-best structure, the triangular lattice structure.
In \cref{fig performance}(a), a mojority part of $g_{ap}$ belongs to the smallest interval, 
which means that the SPQPD method can design the most suitable processor for a quantum algorithm.
\end{small}
\subsubsection*{\begin{small}Using Random Cricuits\end{small}}
\begin{small}
To further verify the effectiveness of SPQPD method, we evaluate its performance to circuits with different qubit numbers and depth.
Random circuits within the desired qubit number and depth range are generated for such evaluation.

Since NISQ processor has dozens to hundreds of qubits, we set the range of qubit numbers to be $[30, 300]$.
In \cite{cross2018}, the order of the coherence time is in $[10,200]\mu\text{s}$ and the gate execution time satisfies $t_\text{gate}>10\text{ns}$.
Based on the data and the noise model in \cite{AQuantumEngineer'sGuidetoSuperconductingQubits}, if we choose $T=100\mu\text{s},t_\text{gate}=20\text{ns}$,
for a circuit with depth $d=2000$, after rough calculation, the fidelity of computation result is less than $70\%$. At this depth, the fidelity is very low,
thus the circuit depth won't exceed a few thousand and we set the range of depth to be $[50,2000]$.

Fig.\;\ref{fig performance2}(a)-(b) are the relationships between $g_{ap}$ between qubit number and circuit depth.
For error bars of \cref{fig performance2}, we sample 100 random circuit outputs. Error bars are obtained by computing the credible intervals for the data set of circuits with the same size.
These intervals are computed with normal distribution, with a credible level of $95\%$. This ensures that the mean value is inside the credible interval with a probability of 
at least $95\%$.

In \cref{fig performance2}(a), if we consider the mean value $\langle g_{ap}\rangle_d$ of every data bar of circuits with depth $d$, 
the structures designed by the SPQPD method has the smallest $\langle g_{ap}\rangle_d$.
To find out the trend of $\langle g_{ap}\rangle_d$ as depth increases, we use the Mann-Kendall trend test \cite{mannkendall}. 
After computation, the statistics $Z_c$ for SPQPD, triangle structure, Li's structure, and cross square structure are -3.92, -0.12, 0.59, and 2.81.
Therefore, at the significance level $\alpha=0.05$, the SPQPD method has an decreasing trend, cross square structure increasing and the other two have no obvious trend.
This result reveals that the SPQPD method has good scalability for depth. 

In \cref{fig performance2}(b), the $g_{ap}$ increases as the qubit number increases.
Using linear regression, the slope of the SPQPD method is the smallest one 0.0105.
The growth rate of SPQPD method is improved by $37.1\%$ compared with the second-best method, Li's method.

In summary, all results show that the SPQPD method can design the most suitable planar quantum processor for quantum algorithm.
It is an effective way to reduce the number of extra gates and improve the computation fidelity.
As the scale of the algorithm expands, the structure designed by the SPQPD method's advantage becomes more obvious over other structures.
\end{small}
\section*{\begin{large}DISCUSSION\end{large}}
\begin{small}
To improve the quantum computation fidelity, instead of following the old paths of optimizing the transformation process or designing quantum processor with lattice-graph structures,
we come up with an idea that designing the quantum processor with general planar graph structures according to the quantum algorithm.
In particular, we formalize our SPQPD method with four steps:
profiling two-qubit gate information, pruning edges of unimportant vertexes, 
handling the high degree vertexes, and recovering the edges with large weights. 
The numerical experiments show that SPQPD method is a competitive alternative approach to improve the computation fidelity when executing quantum algorithms on NISQ processor.
Therefore, the SPQPD method has great potential to demonstrate the quantum advantage in the NISQ era.

In conclusion, this work explores a step in mitigating the quantum software-hardware gap and
provides a new idea for the development of special-purpose quantum processor.
With the development of technology, SPQPD method might be used in the design of larger NISQ processor and multi-layer processor.
Even after fault-tolerant quantum computation is realized, the SPQPD method will still have practical significance.
PCG designed for certain CCG can reduce the number of swap gates and the circuit depth and therefore
result in decreasing the execution time of quantum algorithms.
\end{small}
\section*{\begin{large}METHODS\end{large}}

\subsubsection*{Profiling Subroutine}\label{subsect profile}
\begin{figure*}[htbp]
\begin{minipage}[t]{0.45\textwidth}
\centering
\includegraphics[width=\textwidth]{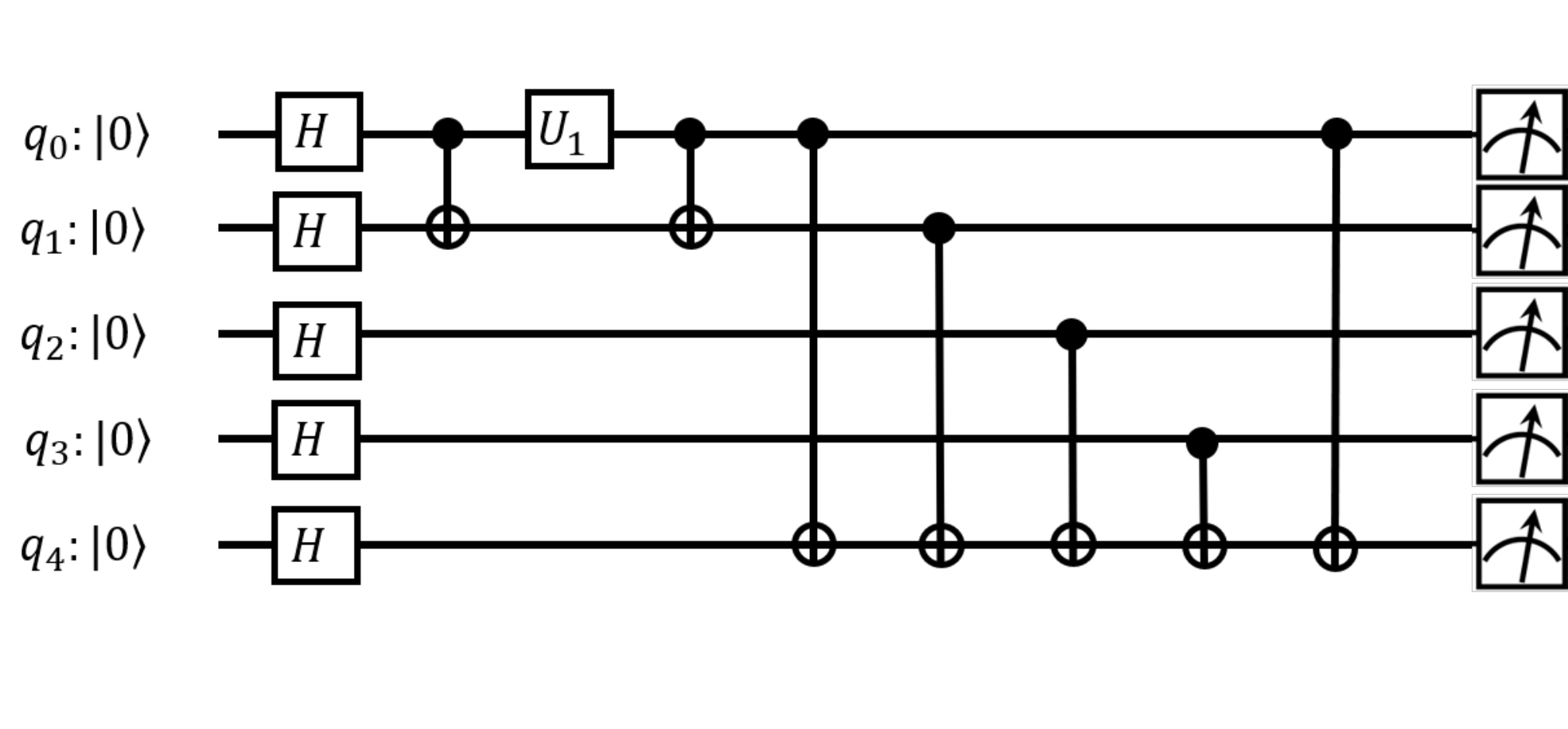}
(a)\label{fig profiledemo1}
\end{minipage}%
\hfill
\begin{minipage}[t]{0.3\textwidth}
\centering
\includegraphics[width=\textwidth]{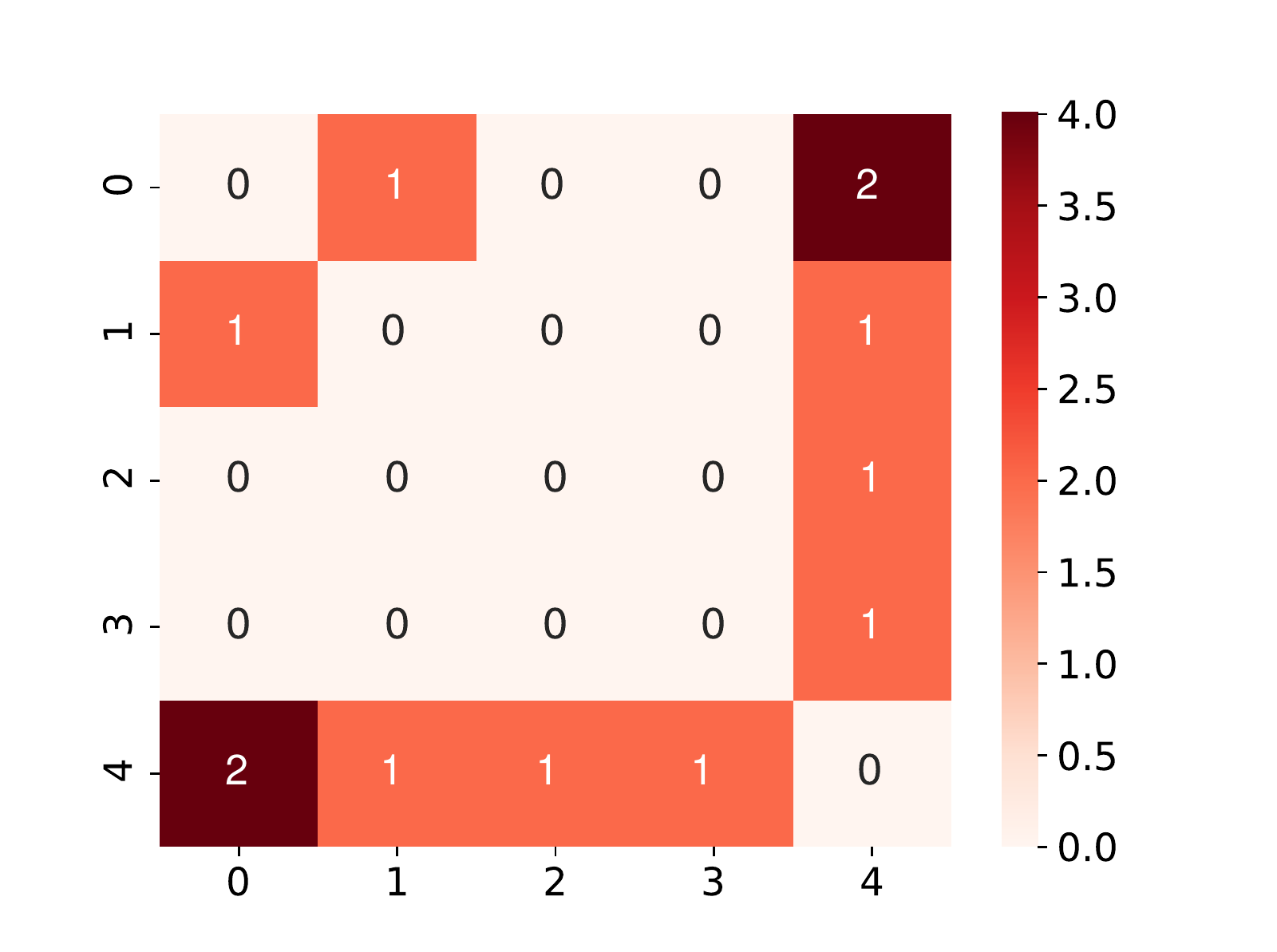}
(b)\label{fig profiledemo2}
\end{minipage}
\hfill
\begin{minipage}[t]{0.2\textwidth}
\centering
\includegraphics[width=\textwidth]{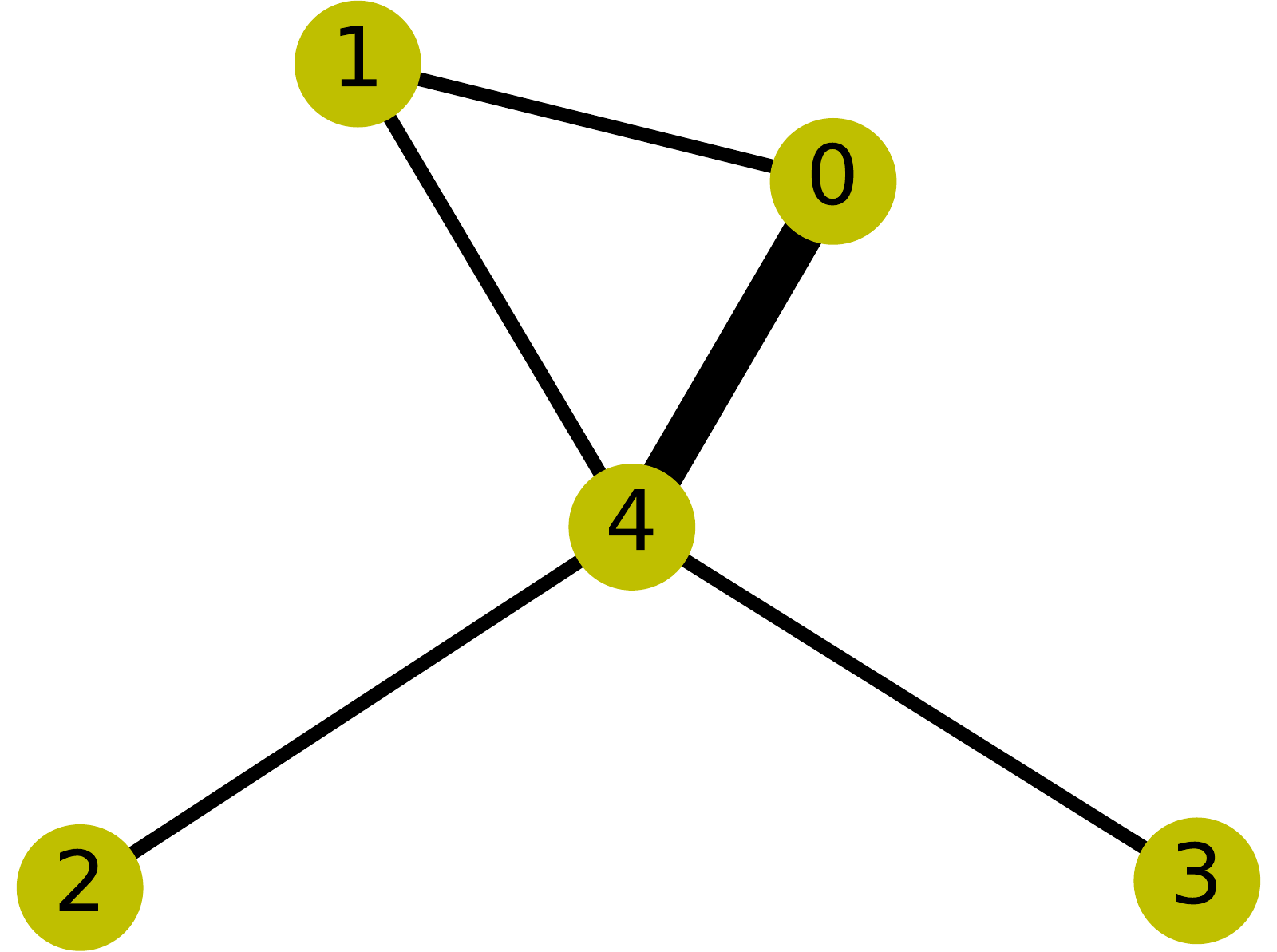}
(c)\label{fig profiledemo3}
\end{minipage}
\caption{\begin{footnotesize}\textbf{Profiling method demonstration.} (a) Original circuit, the two CNOT between $q_0,q_1$ are counted as one two-qubit block. 
(b) The adjacency matrix of (a), whose element $M_{ml}$ is the number of two-qubit blocks between $q_m,q_l$. 
(c) Corresponding CCG of (b), the relative width of the edges represents the relative size of the weight.\end{footnotesize}}\label{fig profiledemo}
\end{figure*}

\begin{small}
We ignore all single-qubit gates, initialization, and measurement operations, and combine the two-qubit gates which act on the same two qubits successively. The reason for
combining is that if the current map allows the first gate to be executed directly, then there is no need for more swap gates for the successive two-qubit gates. 
Therefore, in the circuit transformation process, they can be seen as a whole block. 
After counting the number of two-qubit gate blocks, we get the adjacency matrix $M$ and the corresponding CCG.

Fig.\;\ref{fig profiledemo} shows the process of information profiling. In \cref{fig profiledemo}(a), after ignoring the one-qubit gates and measurement operations, we find that the two CNOT gates
between $q_0$ and $q_1$ can be combined as one two-qubit block. The other CNOTs are two-qubit blocks themselves. Fig.\;\ref{fig profiledemo}(b) is the adjacency matrix of the circuit, whose
element with index $m,l$ represents the number of two-qubit gate blocks between $q_m$ and $q_l$. Fig.\;\ref{fig profiledemo}(c) is the corresponding CCG of \cref{fig profiledemo}(b). 
The relative width of the edges represents the relative size of the weight.
\end{small}
\subsection*{\begin{normalsize}Vertex Ranking Subroutine\end{normalsize}}
\begin{small}
At first, we need to define some concepts about the connection information of the vertexes.
\begin{Def}\label{def pointweight}
The total weight of vertex $q_m$, $W_m=\sum_lM_{ml}$, representing the sum of total two-qubit block number of a vertex.
\end{Def}

\begin{Def}\label{def pointconcentrate}
Weight dispersion of vertex $q_m$, $c_m$, representing the deviation of the number of two-qubit blocks the vertex executes with each of its neighbors.
\end{Def}

Based on the definitions and the discussion above, we choose the following three indicators to evaluate the importance of vertexes.

\textbf{Vertex degree} $D_m$, the larger the degree of the vertex is, the more important the vertex is.

\textbf{Total weight}, according to def.\;\ref{def pointweight}, the larger the total weight, the more frequently this vertex executes two-qubit gates with its neighbors,
and thus more important the vertex is. 

\textbf{Weight dispersion}, the smaller the weight dispersion, the more important this vertex is. The reason why we introduce this indicator will be discussed in the Recovering Subroutine.

Vertexes are sorted by these three indicators in \textbf{lexicographic order}.
\end{small}
\subsection*{\begin{normalsize}Pruning Subroutine\end{normalsize}}
\begin{small}
The first step is choosing the top $N$ most important vertexes as media vertexes. 
The second step is pruning the edges on the graph that don't connect to the media vertexes. 
The third step is storing those pruned edges to a \emph{recover\_set}.
\end{small}
\subsection*{\begin{normalsize}Media Structures Searching Subroutine\end{normalsize}}
\begin{small}
We search for all candidate media structures that are not isomorphism with each other with $n$ vertexes and delete those violate the constraints.  
$n$ is searched from small to large within this range $[2,\infty]$.
If legitimate media structures are found with $n$ vertexes, the searching process stops immediately. 
Because the greater the number of ancilla qubits, the more difficult it is to manufacture the quantum processor.
All of these media structures with $n$ vertexes will be stored and be used as candidate media structures in the following "Weight Allocation Subroutine" subsection. 
\end{small}
\subsection*{\begin{normalsize}Weight Allocation Subroutine\end{normalsize}}
\begin{figure}[h]
\centering
\includegraphics[width=0.4\textwidth]{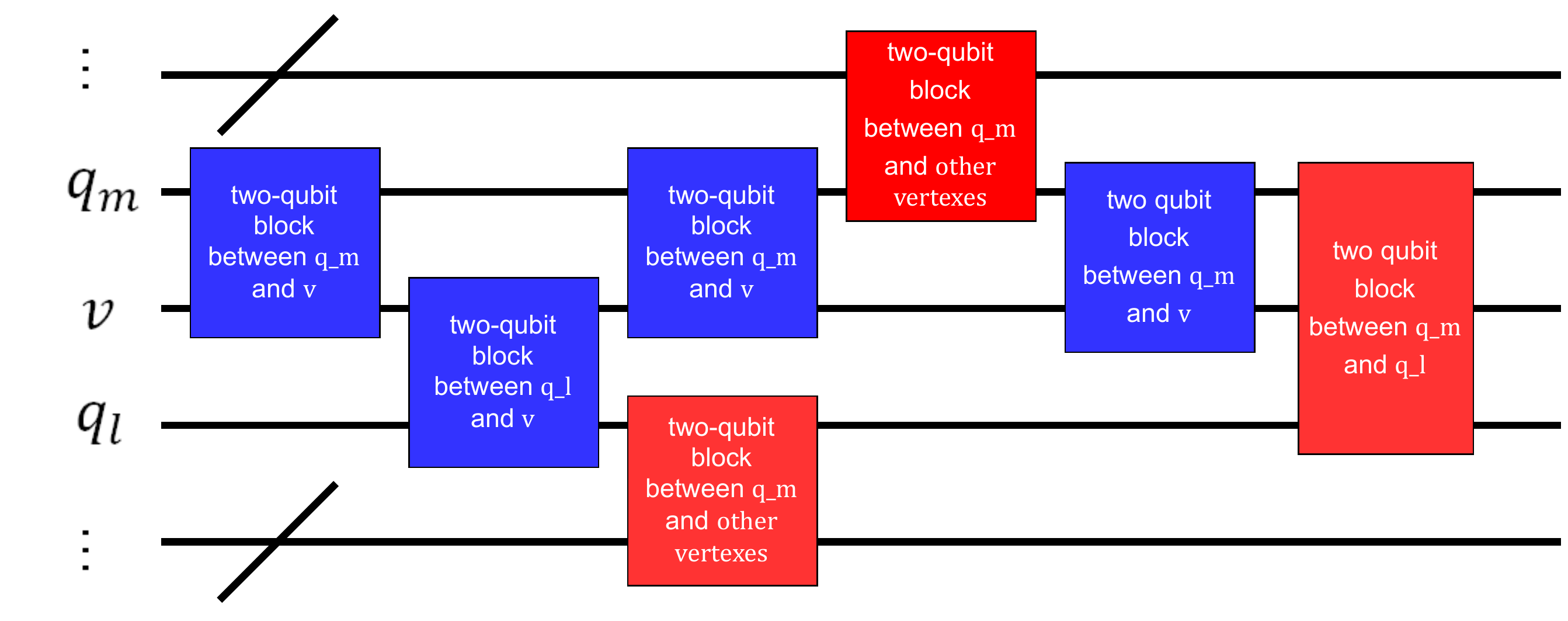}
\caption{\begin{footnotesize}\textbf{Example of the computation of $S_{ml}$.} After ignoring the red blocks, the second and the third blocks between $q_m$ and $v$ can be combined 
as one block. Thus the sum of the two-qubit block number between is $q_m,v$ and $q_l,v$ is 3.\end{footnotesize}}\label{fig sml}
\end{figure}

\begin{small}
First, we compute the element of interaction matrix $I_{ml}=aM_{ml}+(1-a)S_{ml}$.
We give an example \cref{fig sml}(a) of the computation of $S_{ml}$ between $q_m$ and $q_l$. 
When only $q_m,q_l$ and $v$ are of interest, we ignore the operations between other vertexes(the red ones), and combine the successive blocks together.
At last, we get the sum of the number of the two-qubit blocks in the subcircuits of $q_m,v$ and $q_l,v$ which is $S_{ml}=2+1=3$.

Then for every media structure, we compute the element of matrix $C$, whose elements are the shortest length between $v_i,v_j$ in the media structure.

If no non-media vertex has been allocated on the media structure, we computate the indicator, 
\end{small}
\begin{align}
E_m=\sum_lI_{ml},\label{eq m1}
\end{align}
\begin{small} 
\noindent and choose the largest $m$ as the index of the first non-media
vertex to allocate. And the location is the vertex with the largest degree under the degree constraint.

Else if there are non-media vertexes have been allocated, we compute another indicator 
\end{small}
\begin{align}
E_m=\sum_{l\in \text{non-media vertexes have been allocated}}I_{ml}, \label{eq m2}
\end{align} 
\begin{small}
\noindent and choose the largest $m$ as the index of next non-media vertex to allocate.

To find the best location for the non-media vertex $q_m$, we computate 
\end{small}
\begin{align}
s_i=\sum_{j,alloc[j]}C_{ij}I_{alloc[j],m},\label{eq s}
\end{align}
\begin{small}
for every $v_i$ and the $v_i$ with the smallest $s_i$ will be connected to the $q_m$, $alloc[i]\rightarrow alloc[i]\cup q_m$. 
If the degree of $v_i$ break the maximum degree constraint, we repeat the procedure that
excluding $v_i$ and choose the $v_j$ with the second smallest $s_j$ until the constraint is met.

Compute the score according to \cref{eq F}
and choose the media structure with the smallest $F$.
\end{small}
\subsection*{\begin{normalsize}Recovering Subroutine\end{normalsize}}
\begin{figure}[htbp]
\centering
\includegraphics[width=0.4\textwidth]{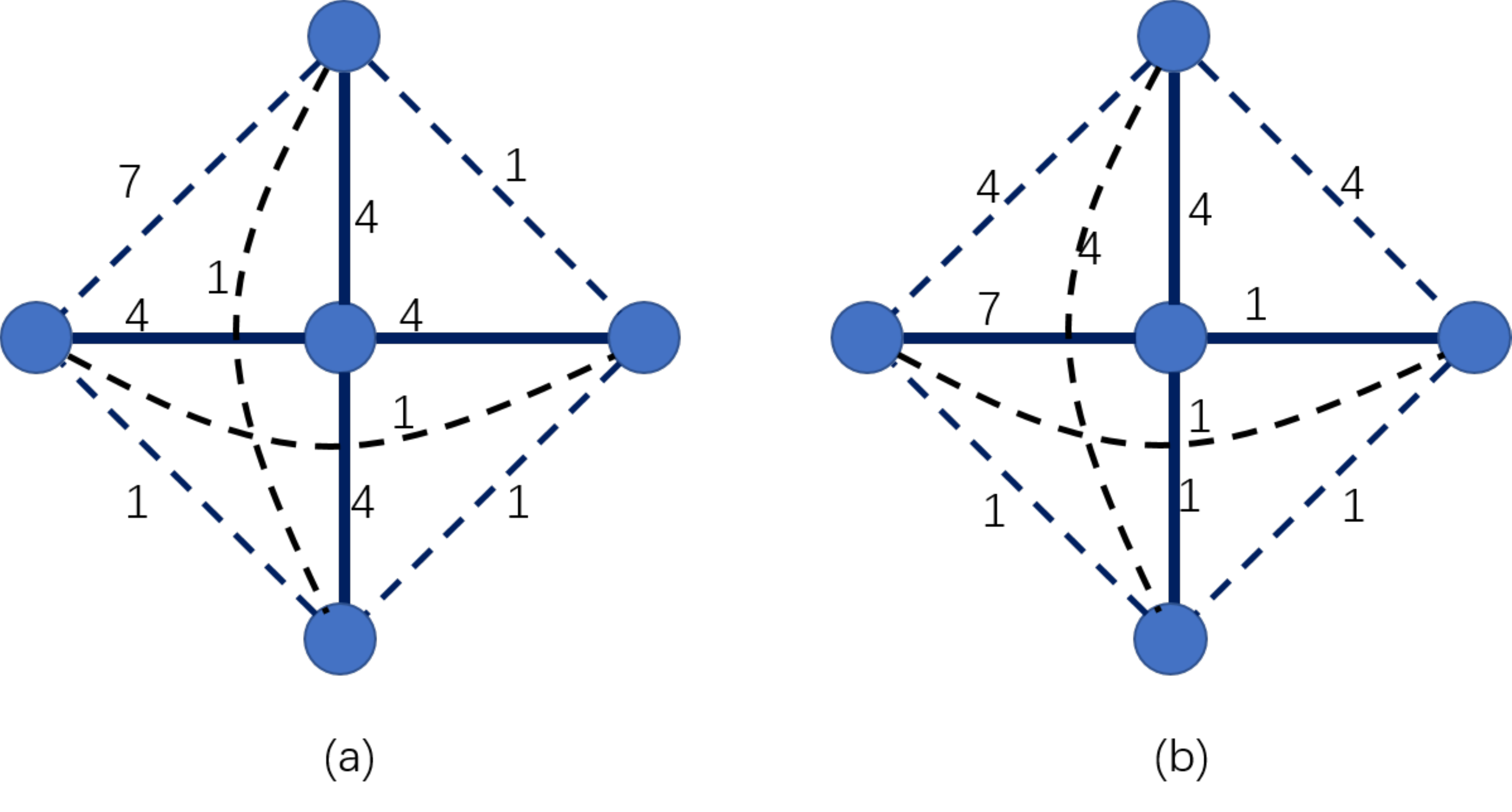}
\caption{\begin{footnotesize}\textbf{Example of recovering of different weight dispersion.} (a) The weight dispersion is smaller. So the pruned edges have larger
dispersion, (b) vice versa. As a result, more weight are concentrated on fewer pruned edges in (a). We can recover a weight 7 by only recovering one edge,
which is impossible in (b).\end{footnotesize}}\label{fig degreevar}
\end{figure}

\begin{small}
We sort the \emph{recover\_set} from large to small by weight.
Then we try to add the edges to the CCG in the order of \emph{recover\_set}.
If the CCG doesn't violate any constraint after edge addition, then updates the CCG, otherwise, delete the edge we add.

The reason that why the vertex with smaller weight dispersion is more important is as follows. 
As in \cref{fig degreevar}, in both cases, the vertexes
in the middle have the same total weight, while the \cref{fig degreevar}(a) has a smaller weight dispersion. As a result, the weight dispersion of edges that are not connected 
to the media vertex is smaller. For instance, more weight is concentrated on the one edge weight 7 in \cref{fig degreevar}(a). While in \cref{fig degreevar}(b), 
the largest weight of edges is only 4, and we can only recover total weight 8 by recovering two edges. 
Consequently, in \cref{fig degreevar}(a), more weight can be recovered by recovering fewer edges.
\end{small}

\subsection*{\begin{normalsize}Determining the Number of Media Vertexes\end{normalsize}}
\begin{small}
We should decide how many high ranking vertexes should be chosen as media vertexes because different circuits have very distinct patterns.
E.g., \cref{fig different CCG}(c), is very different from \cref{fig different CCG}(a). It is reasonable to choose just one media vertex for \cref{fig different CCG}(a) since almost all weights are concentrated on
the edges connected to one vertex. While in \cref{fig different CCG}(c), the weight distribution is uniform. Choosing one media vertex will lose a lot of weight. Therefore, we
try $N$ from 1 to the number of CCG vertexes and compute the $\text{score}_N=\sum_{ml}M_{ml}d_{ml}$, where $d_{ml}$ is the shortest distance between $q_m,q_l$ on the PCG.
The smallest $\text{score}_N$ corresponds to the best choice of the number of media vertexes. 
\end{small}
\subsection*{\begin{normalsize}Experiment setup\end{normalsize}}
\begin{small}
\textbf{Benchmarks}
158 quantum programs are collected from IBM's qiskit \cite{Gadi2016}. These benchmarks cover several important fields and have various sizes for a versatility test of the proposed SPQPD method.

\textbf{Hardware Model for Comparison}
We use cross-square structure, triangular lattice structure and the structures in \cite{li2020}
as representatives of lattice-graph structures and compare them with the SPQPD structure.

\textbf{Circuit Trnasformation Method}
The PCG produced by SPQPD is still different from the original CCG. Thus, optimized transformation process are still required. To illustrate the improvement of our method, 
for all the structures, we use the same Sabre transformation method provided by IBM qiskit \cite{li2019,Gadi2016}.
\end{small}
\subsection*{\begin{normalsize}Simulation Tools\end{normalsize}}
\begin{small}
The simulation and compilation of quantum algorithms are completed by an open-source software development kit (SDK) IBM qiskit \cite{Gadi2016}.
\end{small}

\section*{\begin{small}CODE AND DATA AVAILABLE\end{small}}
\begin{footnotesize}
The experimental data and the code that support the findings of this study are available from the corresponding author on reasonable request.
\end{footnotesize}

\section*{\begin{small}AUTHOR CONTRIBUTIONS\end{small}}
\begin{footnotesize}
B.H.L. designed the method and carried out the numerical calculations. 
B.H.L. and Y.C.W. analyzed the data and wrote the paper with feedback from all authors. 
W.C.K. and Q.Z. gave suggestions on the article writing.
G.P.G. supervised the project.
\end{footnotesize}

\section*{\begin{small}COMPETING INTERESTS\end{small}}
\begin{footnotesize}
The authors declare no competing interests.
\end{footnotesize}

\section*{\begin{small}ADDITIONAL INFORMATION\end{small}}
\begin{footnotesize}
Correspondence and request for materials should be addressed to Y.C.W. or G.P.G..
\end{footnotesize}

\section*{\begin{small}ACKNOWLEDGEMENTS\end{small}}
\begin{footnotesize}
This work was supported by the National Key Research and Development Program of China (Grant No. 2016YFA0301700) and the 
Anhui Initiative in Quantum Information Technologies (Grant No. AHY080000).
\end{footnotesize}
\end{document}